\documentclass[aps,prd,twocolumns,eqsecnum,preprintnumbers,showpacs,amsmath,amssymb]
{revtex4}

\usepackage{graphicx}
\usepackage{dcolumn}
\usepackage{bm}

\newcommand{\dd}{{\textrm d}}
\newcommand{\GeV}{{\textrm{GeV}}}
\newcommand{\MeV}{{\textrm{MeV}}}

\usepackage{floatrow}
\begin{document}

\title{ANALYTIC DESCRIPTION OF $SU(3)$ LATTICE THERMODYNAMICS

IN THE WHOLE TEMPERATURE RANGE WITHIN THE MASS GAP APPROACH}

\vspace{3mm}

\author{V. Gogokhia$^{1,2}$, \ A. Shurgaia$^{2,3}$ \ M. Vas\'uth$^1$}
\email[]{gogohia.vahtang@wigner.mta.hu}
\email[]{gogokhia@rmi.ge}
\email[]{avsh@rmi.ge}
\email[]{vasuth.matyas@wigner.mta.hu}

\vspace{3mm}

\affiliation{$^1$WIGNER RCP, RMI, Depart. Theor. Phys., Budapest 1121,
P.O.B. 49, H-1525, Hungary}

\affiliation{$^2$A. Razmadze Mathematical Inst. of I. Javakhishvili Tbilisi State University,
Depart. Theor. Phys., Tamarashvili str. 6, 0177 Tbilisi, Georgia}

\affiliation{$^3$European School, IB Word School, 34g A. Kazbegi Av., 0177 Tbilisi, Georgia}

\date{\today}
\begin{abstract}
A general approach how to analytically describe and understand $SU(3)$ lattice
thermodynamics in the whole temperature range $[0, \infty)$ is formulated and used. It is based
on the effective potential approach for composite operators properly extended
to non-zero temperature and density. This makes it possible to introduce into this general formalism
the mass gap, which is responsible for the large-scale dynamical structure of the QCD ground state.
The mass gap dependent gluon plasma pressure adjusted by this approach to the corresponding lattice data is shown to
be a continuously growing function of temperature being thus differentiable in every point of its domain.
At the same time, the entropy and energy densities
have finite jump discontinuities at some characteristic temperature $T_c = 266.5 \ \MeV$ with latent heat
$\epsilon_{LH}= 1.41$. This is a firm evidence of the first-order
phase transition in $SU(3)$ pure gluon plasma. The heat capacity has
a $\delta$-type singularity (an essential discontinuity) at $T_c$,
so that the velocity of sound squared becomes zero at this point.
All the independent thermodynamic quantities are exponentially
suppressed below $T_c$ and rather slowly approach their respective
Stefan-Boltzmann limits at high temperatures. Those thermodynamic
quantities which are the ratios of their independent counterparts
such as conformity, conformality and the velocity of sound squared
approach their Stefan-Boltzmann limits rather
rapidly and demonstrate a non-trivial dependence on the
temperature below $T_c$. We also calculate the trace anomaly relation (the interaction measure) and closely related to it the gluon condensate,
which are especially sensitive to the non-perturbative effects. All the calculated thermodynamic quantities have a complicated
and rather different dependence on the mass gap and temperature across $T_c$.
An analytical description of the dynamical structure of $SU(3)$ gluon plasma is given.
\end{abstract}



\maketitle

\section{Introduction}

From the very beginning and up to present days, lattice QCD remains the only practical method to
investigate QCD at finite temperature and density from first principles. Recently it underwent a rapid progress
(see, for example \cite{1,2,3,4,5,6} and references therein). However, lattice QCD is primarily aimed
at obtaining well-defined calculation schemes in order to get
realistic numbers for physical quantities. One may therefore get
numbers and curves for various thermodynamic quantities/observables, but without understanding what is the physics behind
them. Such an understanding can only come from an analytic description of the corresponding lattice data in the whole
temperature range and desirably on a general dynamical ground. So the merger between lattice and analytical approaches to QCD at
finite temperature and density is unavoidable, i.e., they do not exclude each other: on the contrary, they should be complementary.
In other words, numbers and curves come from thermal lattice QCD, while the analytic description of the physics for them comes
from the dynamical theory, which is continuous QCD. There already exist some interesting analytical methods and models based
on the mass gap, quasi-particle, glueball gas, liquid, etc. pictures to analyse results of QCD lattice thermodynamics in different
temperature ranges \cite{7,8,9,10,11,12} (and references therein).

The effective potential approach for composite operators \cite{13,14}
turned out to be a very effective analytical and perspective dynamical tool for the
generalization of QCD to non-zero temperature and density. In the
absence of external sources it is nothing but the vacuum energy
density (VED), i.e., the pressure apart from the sign. This approach is non-perturbative (NP) from the very
beginning, since it deals with the expansion of the corresponding
skeleton vacuum loop diagrams in powers of the Planck constant, and thus allows one to calculate the
VED from first principles. The key element in this program is the
extension of our paper \cite{14} to non-zero temperature \cite{7}.
This makes it possible to introduce the correctly defined temperature-dependent bag
constant (bag pressure) as a function of the mass gap. It is this which is
responsible for the large-scale structure of the QCD ground state \cite{7}.
It coincides with the Jaffe-Witten (JW) mass gap \cite{15} by properties, but not by definition.
The confining
dynamics in the gluon matter (GM) is therefore nontrivially taken
into account directly through the mass gap and via the
temperature-dependent bag constant itself, but other NP effects due to the mass gap are
also present. Being NP, the effective approach for composite operators,
nevertheless, makes it possible to incorporate the thermal
perturbation theory (PT) expansion in a self-consistent way.
In our auxiliary work \cite{16} we have formulated and developed the analytic thermal PT which allows one to
calculate the PT contributions in terms of the convergent series in integer powers of a small
$\alpha_s$. We have explicitly derived and numerically calculated
the first PT correction of the $\alpha_s$-order to the NP part of the GM equation of state (EoS) or, equivalently, the gluon pressure \cite{7,16}.

In this article from the very beginning, we are investigating a system at non-zero temperature, which consists of
$SU(3)$ purely Yang-Mills (YM) gauge fields without quark degrees of freedom (i.e., at zero density).
Within our general approach and for future purpose, it is useful to introduce here a following convention:
by the GM we understand the above-mentioned system at low temperature below
some characteristic temperature $T_c$ ( for its numerical value see section V and Fig. 1). The NP effects are dominant
in this temperature interval. By the gluon plasma (GP) in what follows we will understand the same system at rather high temperature
above $T_c$. However, the NP effects are still important in this region up to moderately high temperature. In the limit of very high
temperature this system will become of the so-called Stefan-Boltzmann (SB) one, i.e, consisting of a free (non-interacting) massless gluons.
Such a separation makes sense, since the system under the consideration is being/staying under rather different extreme conditions
below and above $T_c$ despite the existence of the mixed phase between them (see section IX).

The GM EoS determines the NP thermodynamic structure of the GM at low temperature and still remains important at moderately high
temperature, as underlined above. Due to the initial normalization condition of the
free PT vacuum to zero, the gluon pressure itself cannot achieve the required SB limit at high temperatures.
In order to achieve this limit for the full pressure it is necessary to include into the gluon pressure the above-mentioned SB term.
Since it cannot be simply added to the GM EoS at high temperatures, so it has to be done in a more sophisticated
way (see discussion in section VI). Such obtained full pressure will be called the GP pressure or, equivalently, the GP EoS.
Its low temperature limit is to be determined by the gluon pressure itself, so it will be valid for the whole temperature range.
However, the GP pressure will depend on the two auxiliary functions of temperature, which numerical values can be only fixed by
using the corresponding $SU(3)$ lattice thermodynamics data. How to do this within a self-consistent formalism is precisely one of
the main goals of this article. This will make it possible to analytically describe the above-mentioned lattice thermodynamics
with the help of the mass gap, included
into the framework of the effective potential approach for composite operators generalized to non-zero temperature.
We will explicitly show how to analytically continue the thermal YM lattice calculations near $T_c$ to the region of very low temperatures where they
usually suffer from big uncertainties \cite{1} (and see figures below as well).
One gets an analytic description of all the lattice thermodynamics results on a general dynamical ground and in the whole temperature range.
In other words, the analytic description of the dynamical structure
of $SU(3)$ GP and the quark-gluon plasma (QGP) \cite{1,17,18,19,20,21,22,23} (after inclusion of quarks, i.e., at non-zero density)
will become possible. It will be based on all the possible lattice data for the pressure only,
since all other thermodynamic observables can be analytically expressed and numerically calculated through the pressure.

The present paper is organized as follows. In section II the expression for the gluon pressure as a
function of temperature is present. In section III all the analytical results
for the NP part of the gluon pressure are collected and briefly
explained. In section IV the analytic thermal PT is discussed in
general terms. In section V the gluon pressure up to the first
$\alpha_s$-order contribution is analytically calculated. So in short sections III,
IV and V we describe the results obtained previously in \cite{7,16}.
They are present here for the readers convenience in order to have a general picture at hand.
We also think that it is necessary to do since the gluon pressure discussed in these short sections
determines the NP context of the GP pressure, which is to be analytically and numerically agreed with the lattice thermodynamics
results in the next sections. This is another main subject of the present paper, namely how to generalize our previous results
for a rather short temperature range \cite{7} in order to analytically describe the corresponding lattice data
in the whole temperature range. So in section VI we formulate the procedure how to include
the free massless gluons contribution in a self-consistent way into
the GP pressure. A method of the simulating functions is
proposed and analytical formulae for the numerical simulations are
introduced. This makes it possible to perform analytical and
numerical simulations in order to determine the GP pressure,
satisfying all the thermodynamic limits at low-, close to $T_c$- and high temperature limits. In fact, we formulate a general
method how to analytically describe any lattice thermodynamics results in the whole temperature range,
using the gluon pressure as input, and vice-versa, i.e., it describes how the analytically calculated gluon pressure
is to be changed (especially close to $T_c$ and above it), using the corresponding lattice thermodynamics results for the pressure as input.
In section VII we display and discuss our numerical results for all the thermodynamic quantities/observables,
such as the entropy and energy densities, the heat capacity, etc.,
calculated with the help of the obtained GP pressure. In section VIII the
analytic description of the dynamical structure of the GP is given, and in
section IX we summarize our conclusions. In appendixes A and B the general
expressions for the main thermodynamic quantities as functions of the pressure are given and some analytical formulae
for them are derived, respectively. In appendix C the corresponding $\beta$-function as a function of temperature for
the confining effective charge is fixed. These three appendixes are directly taken from \cite{7}, but we decided to again present them here
explicitly for the readers convenience, since the book \cite{7} itself is not freely available.
In appendix D the solution of the renormalization group
equation for the temperature-dependent PT effective charge is obtained and discussed. In appendix E the latent heat is
analytically and numerically evaluated. Some details of the Least Mean Squares method, which has been
used in our calculations, are present in appendix F.
And finally the results of our numerical calculations of the full GP pressure are shown in Table I.

\section{The gluon pressure at non-zero temperature}

In the imaginary-time formalism \cite{22,23,24}, all of the
four-dimensional integrals can be easily generalized to non-zero
temperature $T$ according to the prescription (let us remind that in \cite{7} and in
this paper the signature is Euclidean from the very beginning)

\begin{equation}
\int {\dd q_0 \over (2\pi)} \rightarrow T \sum_{n=- \infty}^{+
\infty}, \quad \ q^2 = {\bf q}^2 + q^2_0 = {\bf q}^2 +
\omega^2_n = \omega^2 + \omega^2_n, \ \omega_n = 2n \pi T.
\end{equation}
In other words, each integral over $q_0$ of the loop momentum is to be
replaced by the sum over the Matsubara frequencies labeled by
$n$, which obviously assumes the replacement $q_0 \rightarrow
\omega_n= 2n \pi T$ for bosons (gluons).

Introducing the temperature dependence into the gluon pressure \cite{7}, we obtain

\begin{equation}
P_g(T) = P_{NP}(T) + P_{PT}(T) = B_{YM}(T) + P_{YM}(T) + P_{PT}(T),
\end{equation}
where the corresponding terms in frequency-momentum space are:

\begin{equation}
B_{YM}(T) = { 8 \over \pi^2} \int_0^{\omega_{eff}} \dd\omega \
\omega^2 \ T \sum_{n= - \infty}^{+ \infty} \left[ \ln \left( 1 + 3
\alpha^{INP}(\omega^2, \omega^2_n) \right) - {3 \over 4}
\alpha^{INP}(\omega^2, \omega^2_n) \right],
\end{equation}

\begin{equation}
P_{YM}(T) = - { 8 \over \pi^2} \int_0^{\infty} \dd\omega \ \omega^2
\ T \sum_{n= - \infty}^{+ \infty} \left[ \ln \left( 1 + {3 \over 4}
\alpha^{INP}(\omega^2, \omega^2_n) \right) - {3 \over 4}
\alpha^{INP}(\omega^2, \omega^2_n) \right],
\end{equation}

\begin{equation}
P_{PT}(T) = - { 8 \over \pi^2} \int_{\Lambda_{YM}}^{\infty}
\dd\omega \ \omega^2 \ T \sum_{n= - \infty}^{+ \infty} \left[ \ln
\left( 1 + {3 \alpha^{PT}(\omega^2, \omega^2_n) \over 4 + 3
\alpha^{INP}(\omega^2, \omega^2_n)} \right) - {3 \over 4}
\alpha^{PT}(\omega^2, \omega^2_n) \right].
\end{equation}

In frequency-momentum
space the intrinsically non-perturbative (INP) and PT effective charges become

\begin{equation}
\alpha^{INP}(q^2) = { \Delta^2 \over q^2} =
\alpha^{INP}( \omega^2, \omega_n^2) = { \Delta^2 \over \omega^2 + \omega_n^2},
\end{equation}
and

\begin{equation}
\alpha^{PT}(q^2) =  { \alpha_s \over 1 + \alpha_s b_0 \ln ( q^2 / \Lambda^2_{YM})}  =
\alpha^{PT} (\omega^2, \omega_n^2) = { \alpha_s \over 1 + \alpha_s b_0 \ln ( \omega^2
+ \omega_n^2 / \Lambda^2_{YM})},
\end{equation}
respectively. It is also convenient to introduce the following notations:

\begin{equation}
T^{-1} = \beta, \quad \omega = \sqrt{{\bf q}^2},
\end{equation}
where, evidently, in all the expressions ${\bf q}^2$ is the square of
the three-dimensional loop momentum, in complete agreement with the relations (2.1).

In Eq.~(2.6) $\Delta^2$ is the mass gap \cite{7}, mentioned above, which is responsible for the large-scale
structure of the QCD vacuum, and thus for its INP dynamics. Recently we have shown that confining
effective charge (2.6), and hence its $\beta$-function, is a result
of the summation of the skeleton (i.e., NP) loop diagrams, contributing to the full gluon self-energy in the $q^2 \rightarrow 0$ regime. This summation has been performed within the corresponding equation of motion \cite{7} (and references therein). It has been done without violating the $SU(3)$ color gauge invariance of QCD. In more detail (including the interpretation of Eq.~(2.6) and the explanation of all the notations above) the derivation of the bag constant as a function of the mass gap and its generalization to non-zero temperature has been completed in \cite{14} and \cite{7}, respectively.

The PT effective charge $\alpha^{PT}(q^2)$ (2.7) is the generalization to non-zero temperature of the
renormalization group equation solution, the so-called sum of the main PT logarithms \cite{7,23,25,26,27}.
Here $\Lambda^2_{YM} = 0.09 \ \GeV^2$ \cite{28} is the asymptotic scale parameter for $SU(3)$ YM fields,
and $b_0=(11 / 4 \pi)$ for these fields, while the strong fine-structure constant is
$\alpha_s \equiv \alpha_s(m_Z) = 0.1184$ \cite{29}. In Eq.~(2.7) $q^2$ cannot go below $\Lambda^2_{YM}$, i.e.,
$\Lambda^2_{YM} \leq q^2 \leq \infty$, which has already been
symbolically shown in Eq.~(2.5). It is worth reminding that the separation between effective charges (2.6) and (2.7),
and hence between the terms (2.3)-(2.5), is not only exact but it is unique one as well. It has been done by the subtraction of the PT part
from the full gluon propagator with the respect of the mass gap, see \cite{7}.

The NP pressure $P_{NP}(T) = B_{YM}(T)+ P_{YM}(T)$ and the PT
pressure $P_{PT}(T)$, and hence the gluon pressure $P_g(T)$ (2.2),
are normalized to zero when the interaction is formally switched
off, i.e., letting $\alpha_s = \Delta^2=0$. This means that the
initial normalization condition of the free PT vacuum to zero holds
at non-zero temperature as well.

\section{$P_{NP}(T)$ contribution}

One of the attractive features of the confining
effective charge (2.6) is that it allows an exact summation over the Matsubara
frequencies in the NP pressure $P_{NP}(T)$ given by the sum of the integrals (2.3) and (2.4). Collecting all
analytical results summarized in our previous work \cite{7}, we can write

\begin{equation}
P_{NP}(T) = B_{YM}(T)+ P_{YM}(T) = {6 \over \pi^2} \Delta^2 P_1 (T)
+ {16 \over \pi^2} T [P_2(T) + P_3(T) - P_4(T)],
\end{equation}
and

\begin{equation}
P_1(T) = \int_{\omega_{eff}}^{\infty} \dd \omega {\omega \over e^{\beta\omega} -1},
\end{equation}
while

\begin{eqnarray}
P_2(T) &=& \int_{\omega_{eff}}^{\infty} \dd \omega \ \omega^2
\ln \left( 1- e^{-\beta\omega} \right), \nonumber\\
P_3(T)&=& \int_0^{\omega_{eff}} \dd \omega \ \omega^2 \ln
\left( 1 - e^{- \beta\omega'} \right), \nonumber\\
P_4(T) &=& \int_0^{\infty} \dd \omega \ \omega^2 \ln \left( 1 -
e^{- \beta \bar \omega} \right),
\end{eqnarray}
where $\omega_{eff}=1 \ \GeV$ and the mass gap $\Delta^2= 0.4564 \ \GeV^2$ for $SU(3)$ gauge theory have been fixed in
\cite{7,14}, and this choice has been explained as well. Here let us only remind that $\omega_{eff}$ is a scale separating the low-
and high frequency-momentum regions. Then $\omega'$ and $\bar \omega$ are given by the relations

\begin{equation}
\omega' = \sqrt{\omega^2 + 3 \Delta^2} = \sqrt{\omega^2 +
m'^2_{eff}}, \quad m'_{eff}= \sqrt{3} \Delta = 1.17 \ \GeV,
\end{equation}
and

\begin{equation}
\bar \omega = \sqrt{\omega^2 + {3 \over 4} \Delta^2} =
\sqrt{\omega^2 + \bar m^2_{eff}}, \quad \bar m_{eff}= {\sqrt{3}
\over 2} \Delta = 0.585 \ \GeV,
\end{equation}
respectively. It is worth reminding that in the NP pressure (3.1)
the bag pressure $B_{YM}(T)$ (2.3) is responsible for the formation
of the massive gluonic excitations $\omega'$ (3.4), while the YM
part $P_{YM}(T)$ (2.4) is responsible for the formation of the
massive gluonic excitations $\bar \omega$ (3.5).

The so-called gluon mean number \cite{22} is

\begin{equation}
N_g \equiv N_g(\beta, \omega) = {1 \over e^{\beta\omega} -1},
\end{equation}
where $\beta$ and $\omega$ are defined in Eq.~(2.8). It
appears in the integrals (3.2)-(3.3) and describes the
distribution and correlation of massless gluons in the GM. Replacing
$\omega$ by  $\bar \omega$ and $\omega'$ we can consider the
corresponding gluon mean numbers as describing the distribution and
correlation of the corresponding massive gluonic excitations in the
GM, see integrals $P_3(T)$ and $P_4(T)$ in Eq.~(3.3). They are of NP
dynamical origin, since their masses are due to the mass gap
$\Delta^2$. All three different gluon mean numbers range
continuously from zero to infinity \cite{22}. We have the two
different massless excitations, propagating in accordance with the
integral (3.2) and the first of the integrals (3.3). However, they
are not free, since in the PT $\Delta^2=0$ limit they vanish (the
composition $[P_2(T) + P_3(T) - P_4(T)]$ becomes zero in this case).
So the NP pressure describes the four different gluonic excitations
(see section VIII below as well). The gluon mean
numbers are closely related to the pressure. Its exponential
suppression in the $T \rightarrow 0$ limit and the polynomial
structure in the $T \rightarrow \infty$ limit are determined by the
corresponding asymptotics of the gluon mean numbers. The low- and
high-temperature expansions for the NP pressure (3.1) have been derived in \cite{7,16}.

Concluding, let us emphasize that the effective scale $\omega_{eff}$ is not an independent scale parameter.
From the stationary condition at zero temperature \cite{14} and the scale-setting scheme at non-zero temperature \cite{7} it follows that

\begin{equation}
\omega_{eff}^2 =(0.4564)^{-1} \Delta^2,
\end{equation}
so it is expressed in terms of the initial fundamental scale parameter - the mass gap.
Its introduction is convenient from the technical point of view in order to simplify our expressions which otherwise will be rather cumbersome.

\section{Thermal PT }

Our primary goal in the previous article \cite{16} was to develop the
analytic formalism for the numerical calculation of the PT term (2.5). It makes it possible to calculate the PT
contribution (2.5) to the gluon pressure (2.2) in terms of the
convergent series in integer powers of a small $\alpha_s$. For this
goal, it is convenient to re-write the integral (2.5) as follows:

\begin{equation}
P_{PT}(T) = - { 8 \over \pi^2} \int_{\Lambda_{YM}}^{\infty}
\dd\omega \ \omega^2 \ T \sum_{n= - \infty}^{+ \infty} \left[ \ln [
1 + x(\omega^2, \omega^2_n)] - {3 \over 4} \alpha^{PT}(\omega^2,
\omega^2_n) \right],
\end{equation}
where

\begin{equation}
x(\omega^2, \omega^2_n) = {3 \alpha^{PT}(\omega^2, \omega^2_n) \over 4 + 3
\alpha^{INP}(\omega^2, \omega^2_n)} = {3 \over 4} {( \omega^2 + \omega^2_n) \over M (\bar \omega^2, \omega_n^2)}
{ \alpha_s \over (1 +  \alpha_s \ln z_n)}
\end{equation}
with the help of the expressions (2.6) and (2.7), and where

\begin{equation}
M(\bar \omega^2, \omega^2_n) = \bar \omega^2 + \omega^2_n, \quad
\ln z_n \equiv \ln z (\omega^2,\omega^2_n) = b_0 \ln [(\omega^2 + \omega^2_n)/ \Lambda^2_{YM}],
\end{equation}
and $\bar \omega^2$ is given in Eq.~(3.5). Let us also note that in these notations

\begin{equation}
\alpha^{PT}(\omega^2, \omega^2_n) \equiv \alpha (z_n) = { \alpha_s \over (1 +  \alpha_s \ln z_n)}.
\end{equation}

Collecting all results obtained in \cite{7,16}, we are able to present the PT part of the gluon pressure as a
sum of the two terms, namely

\begin{equation}
P_{PT}(T) = P_{PT}(\Delta^2; T) + O(\alpha_s^2),
\end{equation}
where

\begin{equation}
P_{PT}(\Delta^2; T) =  \sum_{k=1}^{\infty} \alpha_s^k P_k (\Delta^2; T)
\end{equation}
with

\begin{equation}
P_k(\Delta^2; T) = { 9 \over 2 \pi^2} \Delta^2
\int_{\Lambda_{YM}}^{\infty} \dd\omega \ \omega^2 \ T \sum_{n= -
\infty}^{+ \infty} \left[ { 1 \over M(\bar \omega^2, \omega^2_n)}
(-1)^{k-1} \ln^{k-1} z_n \right].
\end{equation}

Here $P_{PT}(\Delta^2; T)$ (4.6) describes the $\Delta^2$-dependent
PT contribution to the NP term $P_{NP}(T)$ (3.1), beginning with the
$\alpha_s$-order term. In fact, the whole expansion (4.6) is the correction in integer powers of $\alpha_s$ to the NP term $P_{NP}(T)$ (3.1),
i.e., to call it the PT term is only convention.
The $\alpha^2_s$-order term is also a sum of the two
terms, one of which depends on the mass gap and the other one does not.
The corresponding convergent expansions in integer powers of a small $\alpha_s$ for them
begins with $\alpha^2_s$-order terms, see \cite{7,16}. They are not shown
explicitly, since numerically they are much smaller than the first term in Eq.~(4.5).
For this reason their consideration will be omitted in what follows.

\section{The gluon pressure $P_g(T)$}

Taking into account the above-mentioned remarks, the gluon pressure (2.2) then becomes

\begin{equation}
P_g(T) = P_{NP}(T) + P^s_{PT}(T).
\end{equation}
Here $P^s_{PT}(T)$ is the $\alpha_s$-order term
in the expansion (4.6) for $P_{PT}(\Delta^2; T)$, namely
$P^s_{PT}(T) = \alpha_s P_1(\Delta^2; T)$, and for $P_1(\Delta^2;
T)$ see Eq.~(4.7). It is better to re-write the NP pressure (3.1) in
a slightly different form, namely $P_{NP}(T) = \Delta^2 T^2 - (6 /
\pi^2) \Delta^2 P'_1 (T) + (16 / \pi^2) T M(T)$, where $M(T)=
[P_2(T) + P_3(T) - P_4(T)]$. In the integral (4.7) for $k=1$ the
summation over the Matsubara frequencies can be performed
analytically (i.e., exactly). For the explicit expressions of the
integrals $P'_1 (T)$ and $P^s_{PT}(T)$ see below.

It is instructive to gather all our results obtained previously \cite{7,16} for the gluon
pressure as follows:

\begin{equation}
P_g(T) = \Delta^2 T^2 - {6 \over \pi^2} \Delta^2 P'_1 (T) + {16 \over \pi^2} T M(T) + P^s_{PT}(T),
\end{equation}
where the integrals $P'_1(T)$ and $P^s_{PT}(T)$ are

\begin{equation}
P'_1(T) = \int^{\omega_{eff}}_0 \dd \omega \ \omega \ N_g(\beta, \omega) =
\int^{\omega_{eff}}_0 \dd \omega {\omega \over e^{\beta\omega} -1},
\end{equation}
and

\begin{equation}
P^s_{PT}(T) \equiv P^s_{PT}(\Delta^2;T) =  \alpha_s \times {9 \over 2 \pi^2} \Delta^2
\int_{\Lambda_{YM}}^{\infty} \dd\omega \ \omega^2 \
{ 1 \over \bar \omega} { 1 \over e^{\beta \bar \omega}  - 1},
\end{equation}
respectively, while all other integrals $P_n(T), \ n=2,3,4$ are
given in Eq.~(3.3). Here it is worth noting only that the PT term
(5.4) describes the same massive gluonic
excitations $\bar \omega$ (3.5), but their propagation, however,
suppressed by the $\alpha_s$-order. We can consider it as a new
massive excitation in the GM, denoted it as $\alpha_s \cdot \bar
\omega$. Let us remind once more that the term $P^s_{PT}(T)$ is NP, depending
on the mass gap $\Delta^2$, which is only suppressed by the $\alpha_s$ order.
When the interaction is formally switched off, i.e.,
letting $\alpha_s= \Delta^2=0$, the above-defined composition $M(T)$ becomes zero,
as it follows from Eqs.~(3.3), and thus the gluon pressure $P_g(T)$
itself shown in (5.2). This is due to the normalization condition of the free PT
vacuum to zero also valid at non-zero $T$, as emphasized above.

The gluon pressure (5.1) or, equivalently, (5.2) has been calculated and discussed in
\cite{7,16}. It is shown in Fig. 1 and its numerical values
are present in Table I, where the numerical values of its components are also shown.
From this Table one can conclude that $P^s_{PT}(T)$ term is one order of magnitude smaller that $P_{NP}(T)$ term
up to moderately high temperature, while in the limit of high temperature it becomes dominant (see discussion at the end
of this section as well). This effect can be explicitly seen in figures shown in \cite{7,16}.

The gluon pressure (5.2) has a maximum at some "characteristic" temperature, $T_c = 266.5 \ \MeV$.
Below $T_c$ the gluon pressure is exponentially suppressed in the $T \rightarrow 0$ limit, namely

\begin{eqnarray}
\hspace{-15mm} P_g(T) &\sim& {6 \over \pi^2} \Delta^2 ( T^2 + \omega_{eff} T) e^{ -  \nu_1{T_c \over T }}
- {16 \over \pi^2} T \left[ 2T^3 + 2 \omega_{eff} T^2 + \omega^2_{eff} T \right] e^{ - \nu_1{T_c \over T }}
\nonumber\\
&+& {16 \over \pi^2} T \left[ ( 2T^3 + 2 \omega'_{eff} T^2 + \omega'^2_{eff} T) e^{ - \nu_2 {T_c \over T }}
-( 2T^3 + 2 \sqrt{3} \Delta   T^2 + 3 \Delta^2 T) e^{ - \nu_3 { T_c \over T }}                                  \right] \nonumber\\
&-& {24 \over \pi^2} T^2 \Delta^2 \left[ e^{ - \nu_2 {T_c \over T }} - e^{ - \nu_3{ T_c \over T }} \right]
+ {16 \over \pi^2} T \left[ 2T^3 + \sqrt{3} \Delta T^2 +
{3 \over 8} \Delta^2 T \right] e^{ - \nu_4 { T_c \over T }} \nonumber\\
&+& {9 \alpha_s \over 2 \pi^2} \Delta^2 \left[ (T^2 + T \tilde{\omega}_{eff}) e^{- \nu_5 {T_c \over T}} + {3 \over 8} \Delta^2 {\rm Ei}( - \nu_5 {T_c \over T}) \right], \quad T \rightarrow 0.
\end{eqnarray}
which is related to the low-temperature asymptotic of the gluon mean number
(3.6), as mentioned above. For the numerical values of the exponents $\nu_i, \ i=1,2,3,4,5$ see our work \cite{16}, where
the low-temperature expansion (5.5) has been analytically derived (for the explicit expressions of
$\omega'_{eff}$, $\bar \omega_{eff}$ and $\tilde{\omega}_{eff}$ see below).
Its characteristic features are: A non-analytical dependence on
the mass gap in terms $\sim (\Delta^2)^{1/2}T^3 \sim \Delta T^3$, though the mass gap $\Delta^2$ is not an expansion parameter like
$\alpha_s$. The presence of terms $\sim T^4$, the so-called SB-type terms, though overall coefficient in front of them vanishes
in the PT $\Delta^2=0$ limit, as it has been derived in \cite{16}.
Close to $T_c$ the expansion for $P_g(T)$ can be obtained from the expression (5.5) by putting $T=T_c - \delta T$
and expanding in powers of a small $\delta= - 1 + (T_c/T)$ in the $T \rightarrow T_c$ limit. So it explicitly shows an exponential rise
of the number of dynamical degrees of freedom in this limit.

At moderately high temperatures up to approximately $(4-5)T_c$ the exact functional
dependence on the mass gap $\Delta^2$ and temperature $T$ of the gluon pressure (5.2)
remains rather complicated. This means that the NP effects due to
the mass gap are still important up to rather high temperature. The gluon
pressure has a polynomial character in integer powers of $T$ up to
$T^2$ at high temperatures. As mentioned above, it is related to the corresponding asymptotic of the gluon mean number
(3.6). Its high-temperature expansion analytically derived in \cite{16} is as follows:

\begin{eqnarray}
\hspace{-15mm} P_g(T) &\sim& {12 \over \pi^2} \Delta^2 \omega_{eff}T  + {8 \over 3 \pi^2} \omega^3_{eff} T \ln \left( { \omega'_{eff}\over \bar \omega_{eff} } \right)^2 \nonumber\\
&+& {2 \sqrt{3} \over \pi^2} \Delta^3 T\arctan \left({ 2 \omega_{eff} \over \sqrt{3} \Delta} \right) - {16 \sqrt{3} \over \pi^2} \Delta^3 T \arctan \left({ \omega_{eff} \over \sqrt{3} \Delta} \right) \nonumber\\
&+& {9 \over 2 \pi^2} \alpha_s \Delta^2 \left[ {\pi^2 \over 6} T^2 -
T \left( \Lambda_{YM} - {\sqrt{3} \over 2} \Delta \arctan \left( {2 \Lambda_{YM} \over \sqrt{3} \Delta} \right) \right) \right], \ T \rightarrow \infty. \nonumber\\
\end{eqnarray}
Here a non-analytical dependence on
the mass gap occurs in terms $\sim (\Delta^2)^{3/2}T \sim \Delta^3 T$, though
the mass gap $\Delta^2$ is not an expansion parameter like $\alpha_s$, as noted above.
The term $\sim T^2$ has been first
introduced and discussed in the phenomenological EoS \cite{30} (see
also \cite{7,16,31,32,33,34,35,36,37,38} and references therein). On the contrary,
in our approach both terms $\sim T^2$ and $\sim T$ have not been
introduced by hand. They naturally appear on a general ground as a result of the
explicit presence of the mass gap from the very beginning in the NP
analytical EoS (5.2). It is interesting to note that the effective massive gluonic "excitations"
$\omega'_{eff} = \sqrt{\omega^2_{eff} + 3 \Delta^2}$ and $\bar \omega_{eff} = \sqrt{\omega^2_{eff} + (3/4) \Delta^2}$ are logarithmically suppressed at high temperatures, while there is no dependence on the effective massive gluonic "excitation"  $\tilde{\omega}_{eff} = \sqrt{\Lambda^2_{YM} + (3/4) \Delta^2}$.

In a more compact form the previous expansion looks like

\begin{equation}
P_g(T) \sim B_2 \alpha_s \Delta^2 T^2 + [B_3 \Delta^3 + M^3]T, \quad T \rightarrow \infty,
\end{equation}
where the first leading term, which analytically depends on the mass
gap $\Delta^2$, comes from the PT part of the gluon pressure (more precisely from the NP part which
is the $\alpha_s$-order suppressed). The second term, which
dependence on the mass gap is not analytical, since
$(\Delta^2)^{3/2}= \Delta^3$, comes from both parts of the gluon
pressure $P_g(T)$, while $M^3$ denotes the terms of the dimensions of the
$\GeV^3$, which depend analytically on the mass gap $\Delta^2$.
The explicit expressions for it and for both constants $B_2$ and $B_3$ can be easily
restored from the expansion (5.6), if necessary.

Concluding, let us note that the first term $\Delta^2 T^2$ in the gluon pressure (5.2) plays
a dominant role in the region of moderately high temperatures approximately up to $(4-5)T_c$. In the limit of very
high temperatures it is exactly cancelled by the term coming from the composition $M(T)$ in Eq.~(5.2), as it has been established in \cite{16},
where one can also find the discussion of asymptotics of $P_g(T)$ in more detail.
In other words, the $\sim \Delta^2 T^2$ behavior of $P_g(T)$ is replaced by $\sim \alpha_s \Delta^2 T^2$ behavior at very high temperature,
as it should be, in principle. It would be very surprised if a pure NP contribution were survived in the limit of very high temperature,
while for its PT counterpart/correction it would be expected/possible. At the same time, the second purely NP term $\sim T$ is
suppressed in comparison with the first PT term in the very high temperature limit in Eq.~(5.7), indeed.

\begin{figure}
\begin{center}
\includegraphics[width=10cm]{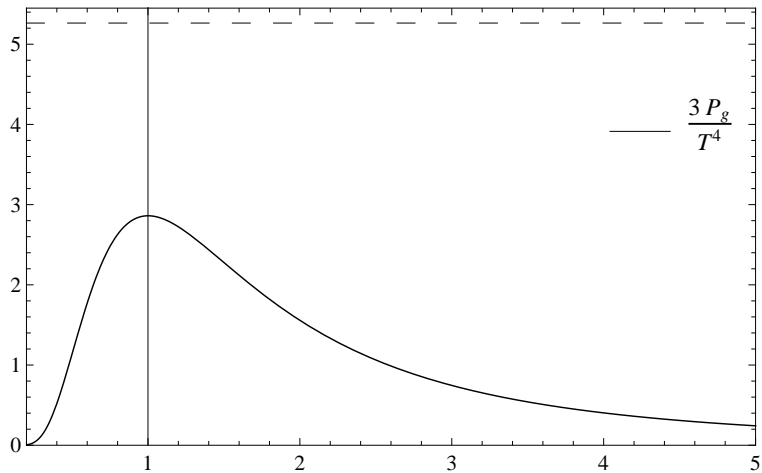}
\caption{The gluon pressure (5.2) scaled (i.e., divided) by $T^4 /3$ is shown as a function of $T/T_c$ (solid curve).
It has a maximum at $T_c = 266.5 \ \MeV$ (vertical solid line). The horizontal dashed line is the general SB constant (A8).}
\label{fig:1}
\end{center}
\end{figure}

\section{The full GP EoS}

From Fig. 1 it clearly follows that the gluon pressure (5.2) will
never reach the general SB constant (A8) at high temperatures. That
is not a surprise, since the SB term has been canceled in the gluon
pressure from the very beginning due to the normalization condition
of the free PT vacuum to zero \cite{7,16}. Analytically this
cancelation at high temperatures (above $T_c$) has been
shown in \cite{16}, where it has also been shown that the massless
(but not free) gluons may be present at low temperatures (below
$T_c$) in the GM. However, their propagation in this region cannot
be described by the SB term itself. All this means that the SB
pressure has been already removed from the gluon pressure, but in
a very specific way, i.e., the above-mentioned normalization
condition is not simply the subtraction of SB term. Let us remind that
the pressure at zero temperature has been normalized as follows: $P \sim \int d^4q \ Tr \left[ \ln (D_0^{-1}D) - (D_0^{-1}D) + 1 \right] $,
where $D$ and $D_0$ are being the full and free gluon propagators, respectively. Just the integral over $D_0$ leads to the SB term
after going to non-zero temperature (see \cite{7,14} for details).

The gluon pressure (5.2) may change its exponential regime below
$T_c$ only in the close neighborhood of $T_c$ in order for its full
counterpart to reach the corresponding SB limit at high
temperatures above $T_c$. The SB term is valid only at high temperatures,
nevertheless, it cannot be added to Eq.~(5.2) above $T_c$, even
multiplied by the corresponding $\Theta ((T / T_c) - 1)$-function.
The problem is that in this case the pressure will get a jump at
$T=T_c$, which is not acceptable. The full pressure is always a
continuous growing function of temperature at any point of its
domain. This means that we should add some other terms valid below
$T_c$ in order to restore a continuous character of the full
pressure across $T_c$. This can be achieved by imposing a special
continuity condition on these terms valid just at $T_c$. Moreover,
the gluon pressure $P_g(T)$ itself should be additionally multiplied
by the functions which are always negative below and above $T_c$. This guarantees
the positivity of the full pressure below $T_c$, while above $T_c$ this guarantees
the approach of the full pressure to the SB limit in
the AF way, i.e., slowly and from below. These terms will also
contribute to the condition of continuity for the full pressure. All
these problems make the inclusion of the SB term into EoS highly
non-trivial. The most general way how this can be done is to add to
Eq.~(5.2) the term $[\Theta((T / T_c) - 1)H(T) + \Theta((T_c/T -
1))L(T)]$, valid in the whole temperature range, and the auxiliary
functions $H(T)$ and $L(T)$ are to be expressed in terms of
$P_{SB}(T)$ and $P_g(T)$ (see subsection A below).

The previous Eq.~(5.2) then becomes

\begin{equation}
P_{GP}(T) = P_g(T) +  \Theta \left( {T_c \over T} - 1 \right) L(T)
+ \Theta \left( {T \over T_c} - 1 \right)H(T),
\end{equation}
and its left-hand side here and below is denoted as $P_{GP}(T)$ (the
above-mentioned full counterpart). Let us emphasize that its low temperature part below $T_c$
was called as the GM pressure above.

The GP pressure (6.1) is continuous at $T_c$ if and only if

\begin{equation}
L(T_c) = H(T_c),
\end{equation}
which can be easily checked. Due to the continuity condition (6.2),
the dependence on the corresponding $\Theta$-functions disappears at
$T_c$, and the GP pressure (6.1) remains continuous  at any point of
its domain. The role of the auxiliary function $L(T)$ is to change
the behavior of $P_{GP}(T)$ from $P_g(T)$ at low (L) temperatures
below $T_c$, especially in its near neighborhood, as well as to take
into account the suppression of the SB-type terms below $T_c$. The
auxiliary function $H(T)$ is aimed to change the behavior of
$P_{GP}(T)$ from $P_g(T)$, as well as to introduce the SB term
itself and its modification due to AF at high (H) temperatures above
and near $T_c$. These changes are necessary, since
in the gluon pressure $P_g(T)$ the SB term is missing
and it cannot be restored in a trivial way. So the
appearance of the corresponding $\Theta$-functions in the GP
pressure (6.1) is inevitable together with the functions $H(T)$ and
$L(T)$, playing only an auxiliary role but still useful from the
technical point of view (see Eq.~(6.2) and appendices B and D).

Concluding, let us emphasize that the gluon pressure $P_g(T)$ (though determined in the whole temperature range), but being the NP part of the
full pressure, is not obliged and cannot reach SB limit at very high temperature (compare Eq.~(5.7) and relations (A8)).
It is the full pressure $P_{GP}(T)$ which is obliged to approach this limit,
and which should be continuously growing function in the whole temperature range from zero to infinity.

\subsection{Analytical simulations}

Actual analytical and numerical simulations - one of the main
subjects of this paper - need to be done in order to reproduce recent $SU(3)$ lattice
thermodynamics results \cite{33,37}. This will make it possible
to fix the NP analytical EoS for the GP (6.1) valid in the whole temperature range.
The space of basic functions, in terms of
which the auxiliary functions $L(T)$ and $H(T)$ should be found, has
already been established. So on the general ground we can put

\begin{eqnarray}
L(T) &=& f_l(T)P_{SB}(T) - \phi_l(T) P_g(T), \nonumber\\
H(T) &=& f_h(T)P_{SB}(T) - \phi_h(T) P_g(T),
\end{eqnarray}
where all the dimensionless functions $f_l(T), \ f_h(T)$ and
$\phi_l(T), \ \phi_h(T)$ will be called simulating functions. We
call the functions $P_g(T)$ and $P_{SB}(T)$ as basic ones, since
they are independent from each other and exactly known. They
determine the structure of the GP pressure (6.1), while the
simulating functions will mainly produce all the necessary
corrections to their corresponding asymptotics and values at $T_c$ (see
below). This also makes it possible to use in what follows the exact
relations, which are resulting from our calculations given in Table I,
namely

\begin{equation}
\Bigl[ P_{SB}(T) - 1.839855 P_g(T) \Bigr]_{T=T_c} = 0, \
\Bigl \{ {\partial \over \partial T} \Bigl[ P_{SB}(T) - 1.839855 P_g(T) \Bigr] \Bigr\}_{T=T_c} = 0.
\end{equation}

Due to the above discussed normalization condition of the free PT vacuum to zero and exponential suppression of $P_g(T)$
in the $T \rightarrow 0$ limit, the contribution which can be measured in terms of $P_{SB}(T)$ may appear below $T_c$, but only if it is
exponentially suppressed (the so-called SB-type term). This has to
be also true for $P_g(T)$, since we need no additional gluon
pressure in the $T \rightarrow 0$ limit. As we already know
(see discussion in the previous sections), we will achieve this goal by choosing the simulating
functions $f_l(T)$ and $\phi_l(T)$ due to the asymptotic of the
corresponding gluon mean number (3.6) in the $T \rightarrow 0 \
(\beta \rightarrow \infty)$ limit. So putting them as functions of
$(T_c/T)$ in the most general form, one obtains

\begin{equation}
f_l(T) = \sum_{i=1}^n A_i e^{- \mu_i (T_c/T)},  \quad \phi_l(T) = \sum_{i=n+1}^m A_i
e^{ - \mu_i (T_c/T)},  \quad m \geq n+1, \quad T \leq T_c,
\end{equation}
where all $\mu_i > 0$ are arbitrary (in other words, we measure positive quantity
$\beta \omega$ in Eq.~(3.6) in terms of $(T_c/T)$ with the help of these numbers). The constants $A_i$ are also arbitrary ones at this stage.

From the GP pressure (6.1) below $T_c$ and relations (6.3) and (6.5) one gets
\begin{eqnarray}
P_{GP}(T) &=& P_g(T) + f_l(T)P_{SB}(T) - \phi_l(T)P_g(T)  \nonumber\\
&\sim& [1 - \sum_{i=n+1}^m A_i e^{- \mu_i(T_c/T)}]P_g(T)+  \sum_{i=1}^n A_i e^{- \mu_i(T_c/T)}P_{SB}(T)
\end{eqnarray}
in the $T \rightarrow 0 \ (\beta \rightarrow \infty)$ limit. The
additional contributions are indeed exponentially suppressed, and
the asymptotic of the GP pressure $P_{GP}(T)$ is mainly determined by the
gluon pressure $P_g(T)$, as it should be. At the same time, the condition

\begin{equation}
1 - \phi_l(T) = 1 - \sum_{i=n+1}^m A_i e^{- \mu_i(T_c/T)} > 0, \quad T \leq T_c,
\end{equation}
should hold in order for the full gluon pressure $P_{GP}(T)$ to approach zero from above.
So this condition together with the condition $f_l(T) > 0$ guarantees that the full pressure $P_{GP}(T)$ will not have zeros below $T_c$.
Evidently, the corresponding gluon mean numbers (6.5) allow to change the value of
the GP pressure from the gluon pressure $P_g(T)$ near $T_c$, as it is expected.

For the simulating function $f_h(T)$ the general choice is
$f_h(T) = 1 - \alpha_s(T)$, where, in accordance with \cite{12,23,37}, we replaced superscript
"PT" by subscript "s" in the notation for $\alpha (T)$.
Evidently, the first term determines the correct SB limit
for the GP pressure (6.1). The second term in this expression mimics
the PT effective charge with AF property as a function of
temperature. Thus the SB limit at high temperatures should be reached in the AF way (see discussion in appendix D).
This should be true for any other independent
thermodynamic quantities, such as the energy and entropy densities,
etc. There are different empirical expressions for $\alpha_s(T)$
\cite{7,12,22,23,37,39,40} (and references therein). Any
such expression can be re-calculated at any given value of $T_c$,
and thus relate the different formulae for $\alpha_s(T)$ to each
other. Here it is convenient to present the expression for $\alpha_s(T)$ as it has been derived
in \cite{12,23}. So repeating the Letessier-Rafelski procedure for $n_f=0$ with $b_0 =11/4 \pi$ and
$b_1 =51/8 \pi^2$ (for details see appendix D), one obtains

\begin{eqnarray}
f_h(T) &=& 1 - \alpha_s(T) = 1 - \left( 0.22037 { 1 \over t} - 0.033 { \ln t \over t^2} \right), \nonumber\\
t &=& 1 + 0.1929 \ln (T/T_c), \quad T \geq T_c = 266.5 \ \MeV.
\end{eqnarray}
This empirical form reproduces the numerical solution for the perturbative $\beta$-function up to the third digit after point. Let us note that the difference between our characteristic temperature shown above, and that of \cite{33,36,37} which is $T_c= 0.629 \sqrt{\sigma} = 264.2 \ \MeV$ for the square root of the string tension  $\sqrt{\sigma} = 420 \ \MeV$ is rather small, since $266.5/264.2 = 1 + 0.0087$. Our characteristic temperature has been exactly determined by the NP part of the full pressure, as described in section V. However, it is worth pointing out once more a very good numerical agreement between these two values, though obtained by completely different analytical and lattice approaches. This good agreement is a strong argument for us to reproduce lattice results near $T_c$ and at $T_c$, of course, since the pressure should be a continuous function across  $T_c$ (see subsections B and C below).
So the function $f_h(T)$ is uniquely fixed (in fact, it is not a simulating one).
Let us note that the numerical values of the coefficients $b_0, \ b_1$ of the renormalization group equation solution for the corresponding $\beta$-function are hidden in the above-shown numbers. For example, $(0.22037/0.1929)= 1.1424...= b_0^{-1}$ up to fourth digit after point, so that the leading contribution to $\alpha_s(T)$ at high temperature $T \gg T_c$ becomes $(b_0 \ln (T/T_c))^{-1}$, indeed.
Evidently, the first term in the expression (6.8) for $\alpha_s(T)$ reproduces the summation of the so-called main PT logarithms. It mimics the expression (2.7) as a function of $T$. In applications at finite temperature, the ratio $(\omega^2 + \omega^2_n)/ \Lambda^2_{YM}$ in Eq.~(2.7) is effectively replaced by the ratio $T/T_c$.
The $\alpha_s$-order term is completely sufficient to calculate first PT correction to the NP part of the full gluon pressure, while its pure PT part will be reproduced more accurately by the term $f_h(T)P_{SB}(T)$, and where the function $f_h(T)$ is explicitly given in Eq.~(6.8).

The simulating function $\phi_h(T)$ has again to be chosen in the
form of the corresponding gluon mean number (3.6), but its
asymptotic has to be taken in the $T \rightarrow \infty \ (\beta
\rightarrow 0)$ limit (see next subsection). It should be a regular
function of $T$ as it goes to infinity in order not to contradict
the asymptotic of $P_g(T)$ in this limit. The asymptotic of the GP
pressure (6.1) at high temperature $T \rightarrow \infty \ (\beta \rightarrow 0)$ thus
becomes

\begin{eqnarray}
P_{GP}(T) &=& P_g(T)+ f_h(T)P_{SB}(T) - \phi_h(T) P_g(T) \nonumber\\
&\sim& [1 - \alpha_s(T)] P_{SB}(T) + [1 - \phi_h(T)] P_g(T),
\end{eqnarray}

where $[1 - \phi_h(T)]$ has to be negative above $T_c$, so that the GP
pressure will approach the term $[1 - \alpha_s(T)] P_{SB}(T)$ from
below at high temperatures, as it is required. At the same time, the
function $\phi_h(T)$ allows to change the value of the GP pressure
(6.1) from the gluon pressure $P_g(T)$ near $T_c$, as it is
expected. Thus we have the general restriction, namely

\begin{equation}
1 - \phi_h(T) < 0, \quad T \geq T_c.
\end{equation}

The explicit expressions for the auxiliary functions $L(T)$ and
$H(T)$ (6.3), via the chosen simulating functions (6.5), are

\begin{equation}
L(T) = \sum_{i=1}^n A_i e^{ -\mu_i(T_c/ T)} P_{SB}(T)  - \sum_{i=n+1}^m A_i e^{- \mu_i(T_c/T)} P_g(T),
\end{equation}
and

\begin{equation}
H(T) = ( 1 -  \alpha_s(T)) P_{SB}(T) - \phi_h(T) P_g(T),
\end{equation}
where $( 1 -  \alpha_s(T))$ is determined by Eq.~(6.8).

The GP pressure (6.1), on account of the relations (6.11) and (6.12), then looks like

\begin{eqnarray}
P_{GP}(T) = P_g(T)  &+&  \Theta \left( {T_c \over T} - 1 \right)
\left[ \sum_{i=1}^n A_i e^{ - \mu_i (T_c/ T)} P_{SB}(T) - \sum_{i=n+1}^m A_i e^{- \mu_i(T_c/T)}  P_g(T) \right]
\nonumber\\
&+& \Theta \left( {T \over T_c} - 1 \right) \left[ ( 1 - \alpha_s(T)) P_{SB}(T)
- \phi_h(T) P_g(T) \right]. \nonumber\\
\end{eqnarray}
From now on we can forget about the auxiliary functions
$L(T), \ H(T)$, though they are still useful from the technical
point of view in the analytical evaluation of the various thermodynamic
quantities (see appendices B, D and E).

From the relations (6.11), (6.12), and using the relations (6.4), it
follows that at $T=T_c$ the relation (6.2) becomes

\begin{equation}
1.839855  \sum_{i=1}^n A_i e^{ - \mu_i} - \sum_{i=n+1}^m A_i e^{- \mu_i} = 1.839855(1 - 0.22037) - \phi_h(T_c)  = 1.4344 - \phi_h(T_c).
\end{equation}

Due to this relation, from the previous expression (6.13) at $T=T_c$, one obtains

\begin{eqnarray}
P_{GP}(T_c) &=& P_g(T_c) +[1.4344 - \phi_h(T_c)] P_g(T_c) \nonumber\\
&=& P_g(T_c) + \left[ 1.839855 \sum_{i=1}^n A_i e^{ - \alpha_i} - \sum_{i=n+1}^m A_i e^{- \mu_i} \right] P_g(T_c),
\end{eqnarray}
which shows that it depends on the number $\phi_h(T_c)$ only, since
the value $P_g(T_c)$ is known from our calculations (see Table I).
From the expression (6.15) it also follows that $2.4344 - \phi_h(T_c) > 0$,
since the full pressure should be always positive, and in particular at $T_c$ it cannot be exactly zero.
Combining now this restriction with the restriction (6.10) at $T=T_c$, one finally obtains

\begin{equation}
1 < \phi_h(T_c) < 2.4344, \quad 1 + 1.839855 \sum_{i=1}^n A_i e^{ - \alpha_i} - \sum_{i=n+1}^m A_i e^{- \mu_i} > 0,
\end{equation}
where the second inequality comes from the second line in the expression (6.15). It is to be compatible with the restriction (6.7)
at $T=T_c$. These relations will be useful for the numerical simulations in what follows.

\subsection{Analytical and numerical simulation of the GP pressure above $T_c$ }

It is convenient to start our procedure of the numerical simulation of lattice results from the region $T \geq T_c$.
Our aim here is to find the simulating function $\phi_h(T)$ by
fitting lattice data in this region. For this, let
us derive from the GP EoS (6.13) its values at $a = T/T_c, \ a \geq 1$ as follows:

\begin{equation}
P_{GP}(a) = f_h(a)  P_{SB}(a) + [ 1 - \phi_h(a)] P_g(a),
\end{equation}
and $f_h(a)$ is given in Eq.~(6.8), namely

\begin{equation}
f_h(a) =  1 - \alpha_s(a) = 1 - \left( { 0.22037 \over (1+ 0.1929 \ln a)} - 0.033 { \ln (1 + 0.1929 \ln a) \over ( 1 + 0.1929 \ln a)^2}
\right),
\end{equation}
with $\alpha_s(1) = \alpha_s(T_c) = 0.22037$ as it follows from Eq.~(6.8) as well.

Adjusting our parametrization of the GP pressure (6.17) to that used
in recent lattice simulations for the YM $SU(3)$ case at $T=aT_c$ in
\cite{33}, one obtains

\begin{equation}
{3 P_{GP}(T) \over T^4 } = {P_l (T) \over T^4 } \times (SB),
\end{equation}
where $(SB) = 3P_{SB}(T) / T^4 = (24/45) \pi^2$ is the general SB constant, see Eq.~(A8), where its numerical value is shown up to fourth digit after point, already properly rounded off. In our numerical calculations we will use its exact value, of course, as well as values of lattice data \cite{33} and our data presented in Table I. All other numbers will be properly round off if possible, for convenience. At the same time, let us stress that we can calculate all our numbers to any requested accuracy. The subscript "l" in $P_l(T)/T^4$ is due
to the above-mentioned lattice data, which, for example should read
$P_l (T_c) / T_c^4 = 0.019676, \ P_l (2T_c) / (2T_c)^4 = 0.613278, \ P_l (3T_c) / (3T_c)^4 = 0.731751$ and so on.
Our values for $P_g(T)$ will be also used with the same accuracy, Table I. However, let us remind that our value for $T_c$
coincides with its lattice counterpart up to third digit after point (see subsection A above). This means that, in principle,
in our numerical calculations we are responsible up to this order only, though we will show numbers beyond it as well.

As we already know, the best way to choose the appropriate
expression for the simulating function $\phi_h(a)$ is to mimic
the asymptotic of the gluon mean number (3.6) but in the $T
\rightarrow \infty \ (\beta \rightarrow 0)$ limit, which is
equivalent to the $a \rightarrow \infty$ limit.  As a function of
$a$, one can write

\begin{equation}
\phi_h(a) = { a^{-1} \over e^{(\mu/a)} -1} = { a^{-1} \over \sum_{k=1}^{\infty} (1 / k!) (\mu/a)^k }
= { 1 \over \sum_{k=0}^{\infty} c'_{k+1} a^{-k}}  = \sum_{k=0}^{\infty} c_k a^{-k},
\end{equation}
where $c'_k = (1 / k!) \mu^k$ and $c_0 c'_1=1, \ c_n + (1 / c'_1)
\sum_{k=1}^n c_{n-k}c'_{k+1} = 0, \ n=1,2,3...$ . So this simulating
function at high temperatures becomes a series in inverse powers of
$a = T /T_c$, starting from non-zero $c_0= \mu^{-1} > 1 $ in
agreement with the general restriction (6.10). This was the reason
for the multiplication of the corresponding gluon mean number in
Eq.~(6.20) by $a^{-1}$. A possible arbitrary constant to which the initial $a^{-1}$ has
to be multiplied is set to one without loosing generality due to the arbitrariness of the constants $c_k$ in the initial expansion (6.20) at this stage.

First of all, we are interested in

\begin{equation}
\phi_h(1) = \sum_{k=0}^{\infty} c_k,
\end{equation}
as it follows from the previous Eq.~(6.20). This important quantity
appears in Eqs.~(6.14) and (6.15), since $\phi_h(1) \equiv
\phi_h(T_c)$. On the other hand, the series (6.20) can be re-written
as follows:

\begin{eqnarray}
\phi_h(a) &=& \sum_{k=0}^{\infty} c_k a^{-k} = \sum_{k=0}^m c_k
a^{-k} + \sum_{k=m+1}^{\infty}
c_k a^{-k} \nonumber\\
&=& \nu_0(a; m) + c_{m+1} a^{-m-1} + \sum_{k=m+2}^{\infty} c_k
a^{-k},
\end{eqnarray}
where

\begin{equation}
\nu_0(a; m) = \sum_{k=0}^m c_k a^{-k} = c_0 + O(a^{-1}), \quad a \rightarrow \infty.
\end{equation}
At $a = 1$ Eq.(6.22) yields

\begin{equation}
c_{m+1} = [\phi_h(1) - \nu_0(1; m)]  - \sum_{k=m+2}^{\infty} c_k .
\end{equation}
Substituting it back to the previous Eq.~(6.22), one obtains

\begin{equation}
\phi_h(a) = \nu_0(a; m) + [\phi_h(1) - \nu_0(1; m)] a^{-m-1} -
a^{-m-1} \sum_{k=m+2}^{\infty} c_k  + \sum_{k=m+2}^{\infty} c_k
a^{-k}.
\end{equation}

This means that the whole expansion (6.20) or, equivalently, (6.25)
can be effectively correctly replaced (approximated) by the two terms
polynomial as follows:

\begin{equation}
\phi_h(a) = c_0 + [\phi_h(1) - c_0] a^{-m-1},
\end{equation}
which has the correct limit when $a$ goes to infinity, Eq.~(6.23),
and it is self-consistent at $a=1$. In other words, in both limits
$a = [1, \infty)$ it behaves like the initial infinite series
(6.25). For future purpose it is convenient to present this equation
in the equivalent form, namely

\begin{equation}
\phi_h(a) = c_0 + [\phi_h(1) - c_0] a^{-n} = 1 + \nu_0 + \nu a^{-n},
\end{equation}
where we have put $n=m+1= 1,2,3....$, $ c_0 = \nu_0 + 1$ and $\nu = [\phi_h(1) - c_0]$, so that
$\phi_h(1) = 1 + \nu_0 + \nu$.
Let us underline that all the three independent parameters $\nu_0$, $\nu$ and $n$ remain arbitrary at this stage.

From the relation (6.16) it follows the restriction, namely
$\phi_h(1) = \phi_h(T_c) < 2.4344$. From the relation (6.10) we
also know that $1 - c_0 < [\phi_h(1) - c_0] a^{-n}$ at any $a$. So
at $a=1$ then it follows that $1 < \phi_h(1)$ in complete agreement with (6.16).
When $a$ goes to infinity this will be guaranteed if $1 - c_0 < 0$ or,
equivalently, $c_0 > 1$ itself. It is convenient to present both
restrictions together as follows:

\begin{equation}
1 < c_0, \quad 1 < \phi_h(1) < 2.4344,
\end{equation}
or, equivalently,

\begin{equation}
0 < \nu_0, \quad 0 < \nu, \quad 0 < \nu_0 + \nu < 1.4344.
\end{equation}

The fit to lattice data only available from the moderately high temperature interval
in \cite{33}, namely $a=1 - 3.436657$ is to be performed with the help of the  following
equation

\begin{equation}
{P_l (T) \over T^4 } \times (SB) =  f_h(a)(SB) + [ 1 - \phi_h(a)] {3 P_g(T) \over T^4}
= f_h(a)(SB) - [ \nu_0 + \nu a^{-n}] {3 P_g(T) \over T^4},
\end{equation}
where $T=aT_c$ and the values of $P_l (T) / T^4$ have been taken
from the used lattice data, while the values of $3 P_g(T)/ T^4$ have been taken from our data (Table I).
$f_h(a)$ is given in Eq.~(6.18) and the relation (6.27) has been also used.

However, it is instructive to find explicitly the relation between the parameters $\nu_0$ and $\nu$ from the very beginning. Evidently, this is possible to do by adjusting the right-hand-side of Eq.~(6.30) to its left-hand-side at
$T=T_c$ or, equivalently, $a=1$. Then from the previous equation, one obtains

\begin{equation}
{P_l (T_c) \over T^4_c } \times (SB) = f_h(1)(SB) - [ \nu_0 + \nu] {3 P_g(T_c) \over T^4_c},
\end{equation}
and substituting the numerical values $3 P_g(T_c) / T_c^4= 2.86098$ taken from our calculations
and $(P_l(T_c) / T_c^4) \times (24 / 45) \pi^2  = 0.019676 \times (24 / 45) \pi^2 = 0.103570$ taken from the above-mentioned lattice data, as well as $f_h(1) = 1- 0.22037 = 0.77963$, one arrives at

\begin{equation}
\nu_0 + \nu = 1.3982,
\end{equation}
which definitely satisfies the restrictions (6.29). Taking into account this relation, Eq.~(6.30) becomes

\begin{equation}
{P_l (T) \over T^4 } \times (SB) = f_h(a)(SB) - [ 1.3982 - \nu (1 - a^{-n})] {3 P_g(T) \over T^4}.
\end{equation}
The best fit has been achieved at

\begin{equation}
\nu_0 = 0.55, \quad \nu = 0.8482, \quad n= 3,
\end{equation}
by using the Least Mean Squares (LMS) method \cite{41}.
According to this method the solution for
these parameters $\nu_0, \ \nu$ and $n$ is a unique one, satisfying to
the general restrictions (6.29). It is worth emphasizing that the average deviation is
minimal at the values (6.34). Details of our calculations are briefly described in appendix F.

Hence the relation (6.27) becomes

\begin{equation}
\phi_h(T) = \phi_h(a) = 1.55 + 0.8482 a^{-3} = 1.55 + 0.8482 (T_c/T)^3, \quad
\phi_h(T_c) = \phi_h(1) = 1 + \nu_0 + \nu = 2.3982,
\end{equation}
determining this function up to the leading and next-to-leading
orders in the $T \rightarrow \infty$ limit.

Thus our method makes it possible to establish the behavior of the GP pressure
in the whole high temperature range $a \geq 1$, reproducing lattice data from the finite interval $a=[1, 3.436657]$ only.
Analytically this equation looks like

\begin{equation}
{3P_{GP} (T) \over T^4 } =  f_h(T)(SB) - [ 0.55 + 0.8482 (T_c/T)^3 ] {3 P_g(T) \over T^4},
\end{equation}
where  $f_h(T)$ is given in Eq.~(6.18). The comparison of analytical curve (6.36) with lattice one \cite{33} is shown in Fig. 2.

\begin{figure}
\begin{center}
\includegraphics[width=10cm]{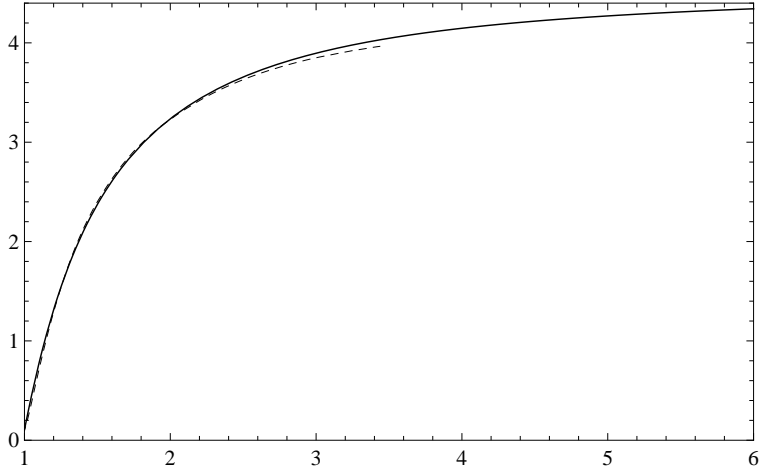}
\caption{The GP pressure (6.36) (solid curve) and lattice $SU(3)$ pressure \cite{33} (dashed curve) are shown as functions of $a \geq 1$. Both curves are calculated in the same units (6.19).}
\label{fig:1}
\end{center}
\end{figure}

\subsection{Analytical and numerical simulation of the GP pressure below $T_c$ }

Our aim here is to find parameters $A_i$ and $\mu_i$ by
fitting lattice data at low temperatures below $T_c$. For this, let
us derive from the GP EoS (6.13) its values at $T_c \geq T$ as follows:

\begin{equation}
P_{GP}(T) =
\sum_{i=1}^n A_i e^{ - \mu_i (T_c/ T)} P_{SB}(T) + [ 1 - \sum_{i=n+1}^m A_i e^{- \mu_i(T_c/T)} ] P_g(T).
\end{equation}

For future purpose it is convenient to continue this subsection with
showing explicitly that the relation (6.14), on account of the numerical value
(6.35) for $\phi_h(T_c)$, becomes equivalent to

\begin{equation}
\sum_{i=1}^n A_i e^{- \mu_i} = 0.543521 \sum_{i=n+1}^m A_i e^{- \mu_i} - 0.523845.
\end{equation}
This relation makes it possible to reduce the number of independent parameters below $T_c$ by one (evidently, it is in agreement with the
second of restrictions in Eq.~(6.16)).
Our aim is to describe the lattice data for the pressure close to $T_c$ by as possible small number of the above mentioned independent
parameters $ A_i, \ \mu_i$. So we started with $n=1, m=2$, i.e., by the one term in each sum over $i$, but we have failed. So the next step
was to start with $n=2, m=3$, i.e., by the two terms from the first sum and again by the one term from the second sum. Moreover, the
constant $A_3$ in the relation (6.5) for $\phi_l(T)$ is put to one without loosing generality and we denote $\mu_3 = \mu$, for simplicity.
We can do this because we need
only suppression of the additional $P_g(T)$ in the $T \rightarrow 0$
limit, while at $T=T_c$ its additional
contribution can be controlled by the parameter $\mu$ only. The relation (6.38) in this case becomes

\begin{equation}
\sum_{i=1}^2 A_i e^{- \mu_i} = 0.543521 e^{- \mu} - 0.523845,
\end{equation}
hinting that $\mu$ should be rather small number, namely $\mu < 0.036872$, assuming that
the sum over $i$ in this relation is always positive, while the lattice pressure at $T_c$ is rather small. In fact, it is not a varying parameter (see below).

The fit to lattice data available from the low temperature interval
in \cite{33}, namely $a= [0.907850 - 1]$ has been performed using the following equation, namely

\begin{eqnarray}
{P_l (T) \over T^4 } \times (SB) &=& \left[ (0.543521 e^{- \mu}
- 0.523845) e^{ \mu_1(1 - (T_c/ T))} + A_2 e^{-\mu_2}[ e^{ \mu_2(1- (T_c/ T))} - e^{ \mu_1(1 - (T_c/ T))}] \right](SB)
\nonumber\\
&+& [1 - e^{ - \mu (T_c/ T)}] {3 P_g(T) \over T^4},
\end{eqnarray}
where the relation (6.39) has been already substituted into Eq.~(6.37). At $T=T_c$, i.e., $a=1$, it is identically satisfied as it should be. The best fit
to lattice data very close to $T_c$ (for the pressure uncertainties of lattice calculations very close to $T_c$ are much smaller than away from it) has been achieved at

\begin{equation}
\mu =0.001, \quad \mu_1 = 39.1, \ \mu_2 = 3.4, \ A_1 e^{-\mu_1} = 0.015732, \ A_2 e^{-\mu_2} =0.003884,
\end{equation}
and, evidently, the chosen value for $\mu$ makes the left-hand-side of the relation (6.39) positive, indeed.
Moreover, any slight deviation from this value changes only the forth digit after point in the numerical value of
the relation (6.39). Let us remind that in our numerical calculations we are responsible
up to third digit after point, as it was underlined in the previous subsection.
Thus, we vary only three independent parameters, $\mu_1, \ \mu_2$ and  $A_2$ in Eq.~(6.40).

Our method makes it possible to establish the behavior of the GP pressure in the whole low temperature range
$a \leq 1$, reproducing lattice data very close to $T_c$ only. Analytically this equation looks like

\begin{equation}
{3P_{GP} (T) \over T^4 } = (SB) \sum_{i=1}^2 A_i e^{ - \mu_i (T_c/ T)} + [1 - e^{ - \mu (T_c/ T)}] {3 P_g(T) \over T^4},
\end{equation}
where the numerical values of all the parameters involved are given in the relations (6.41).  The comparison of analytical (6.42)
curve with lattice \cite{33} one is shown in Fig. 3. For convenience, this fit is shown in Fig. 3 up to $0.2T_c$ only.
May be the fit can be slightly improved by taking into account more
terms in the left-hand-side of the relation (6.38). We restrict ourselves to the two terms only,
since the values of the lattice pressure below $T_c$ are very small and there are no convincing lattice data points below $0.9T_c$ as it follows from Fig. 3. Let us also note that we do not use the LMS method here, since we
have encountered some numerical problems with its non-linear realization. However, our fit made by hand is very accurate for the interval starting
from $0.975T_c$ (i.e., very close to $T_c$, see Fig. 3 again), where we have to trust lattice data, since we believe that they correctly
reproduce the value of the gluon plasma pressure at $T_c$ (see previous subsection).

\begin{figure}
\begin{center}
\includegraphics[width=10cm]{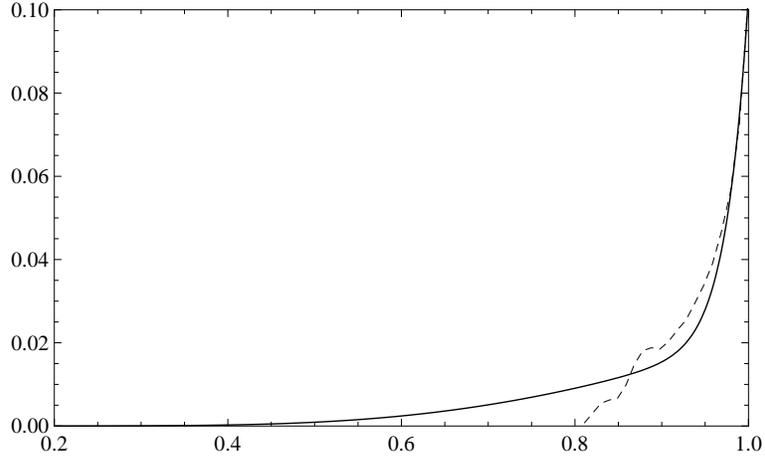}
\caption{The GP pressure (6.42) as a function of $a \leq 1$ is shown
as a solid curve. The lattice curve for $SU(3)$ pressure \cite{33}
is also shown (dashed curve). Very close to $T_c$, starting from $0.975T_c$, they coincide. Both curves are calculated in the same units (6.19).}
\label{fig:1}
\end{center}
\end{figure}

\subsection{The GP pressure in the whole temperature range}

The GP pressure (6.13) in the whole temperature range $T =[0, \infty)$ finally becomes

\begin{eqnarray}
P_{GP}(T) &=& P_g(T) \nonumber\\
&+&  \Theta \left( {T_c \over T} - 1 \right)
\left[ ( 0.015732 e^{- \mu_1 ((T_c/T)-1)} + 0.003884 e^{- \mu_2 ((T_c/T)-1)})
P_{SB}(T) - e^{- \mu (T_c/T)} P_g(T) \right] \nonumber\\
&+& \Theta \left( {T \over T_c} - 1 \right) \left[ (1 - \alpha_s(T) )P_{SB}(T)
- \phi_h(T) P_g(T) \right],
\end{eqnarray}
where

\begin{equation}
\mu =0.001, \quad \mu_1 = 39.1, \ \mu_2 = 3.4,
\end{equation}

\begin{equation}
\alpha_s(T) = \left( 0.22037 { 1 \over t} - 0.033 { \ln t \over t^2} \right), \quad
t = 1 + 0.1929 \ln (T/T_c), \quad T \geq T_c = 266.5 \ \MeV.
\end{equation}

and

\begin{equation}
\phi_h(T) = 1.55 + 0.8482 \left( { T_c \over T} \right)^3, \quad \phi_h(T_c) = 2.3982.
\end{equation}

The GP pressure (6.43) is completely known now, since $P_g(T)$ is
also exactly known, Eq.~(5.2). It is shown in Fig. 4.
This means that the auxiliary functions $L(T)$ and $H(T)$ in
Eq.~(6.1) have finally been fixed in terms of the basic functions
$P_{SB}(T)$ and $P_g(T)$ within our approach (appendix B). For simplicity, in what follows we will omit
the subscript "GP" in the GP pressure (6.43), i.e., we will put $P_{GP}(T) \equiv P(T)$.
The same will be done in the notations of all other thermodynamic quantities as well.

The request that the pressure is
growing continuously as a function of
temperature in the whole range is a rather strong restriction. This
means that it is differentiable at any point of its domain, while at
$T_c$ the derivative itself may not be continuous, i.e., it may have a
discontinuity at this point. At the same time, the adjustment of
both terms at $T_c$, associated with the corresponding
$\Theta$-functions, has to be done in the above-mentioned
requested way, i.e., the pressure should be a continuous function of
the temperature across a possible phase transition at $T_c$ (see
Figs. 4 and 5). There are no other general constraints on the
parameters $\mu_i, \ A_i$ and $\mu$ apart from that the GP pressure (6.43) and
its derivatives should not gain negative
values at low temperatures below $T_c$, i.e., they should
exponentially approach zero from above. Evidently, this will be achieved if we avoid zeros below $T_c$
in the GP pressure itself by fitting the above-mentioned parameters in accordance with lattice data very close to $T_c$.

Our problem was how to restore the free massless gluons
contribution to the full pressure $P_{GP}(T)$ (6.1), maintaining its
continuous character across $T_c$. That is why we use only one type
of the gluon mean numbers for each simulating function, namely the
low-temperature asymptotic of their free massless type (4.6) in the
form of the sum with different $\mu_i$ and $A_i$ parameters,
Eq.(6.5). It is the general one for the simulating function
$f_l(T)$, while for the simulating function $\phi_l(T)$ the chosen
expression is fully sufficient, as explained above in subsection C.
So the choice of the functional form of the simulating functions $f_l(T)$ and $\phi_l(T)$
by their respective expressions is completely justified.
Let us also note that the slight change in the numerical values for the parameters
$\mu_i$ in the relations (6.44) practically nothing changes in the behavior of the pressure (6.43).

For the high-temperature asymptotic the resulting sum (6.20) for the
simulating function $\phi_h(T)$ is the general one, even
using the sum with different parameters. On the one hand, this
makes it possible to achieve the above-mentioned goal. On the other
hand, such choice do not distort the NP content of $P_g(T)$ itself in the whole temperature range.
In other words, we
need the simulating function $\phi_h(T)$ in order for the GP pressure and all its derivatives
to approach their respective SB limits at high temperatures from below.
Let us remind that the simulating function $f_h(T)$ is empirically fixed as a solution of the
corresponding renormalization equation for the PT effective charge (see above and appendix D).
It is needed to ensure the correct SB limit of the GP pressure and its derivatives.

The simulating functions $f_l(T)$ and $\phi_l(T), \ \phi_h(T)$
are needed in order to ensure the continuous character of the GP pressure across $T_c$,
while all the non-trivial PT and NP physics in the GP pressure is due to the basic function $P_{SB}(T)$
and $P_g(T)$, respectively. Concluding, it is worth emphasising once more that there is no other choice for the functional dependence
of these simulating functions as corresponding asymptotics of the gluon mean number. In other words, their functional form is fixed,
but to some finite number of free parameters, as it was described above.

\begin{figure}
\begin{center}
\includegraphics[width=10cm]{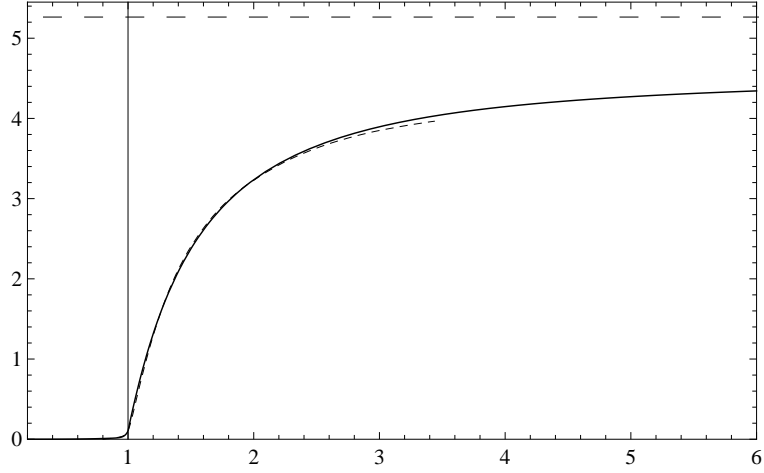}
\caption{The GP pressure (6.43) is shown as a function of $a=T/T_c$
(solid curve). The lattice curve \cite{33} for $SU(3)$ pressure is
also shown (dashed curve). The horizontal dashed line is the general SB constant (A8).
Both pressures are scaled in the same way, see Eq.~(6.19). }
\label{fig:1}
\end{center}
\end{figure}

\section{Numerical results and discussion}

\subsection{GP pressure}

\begin{figure}
\begin{center}
\includegraphics[width=10cm]{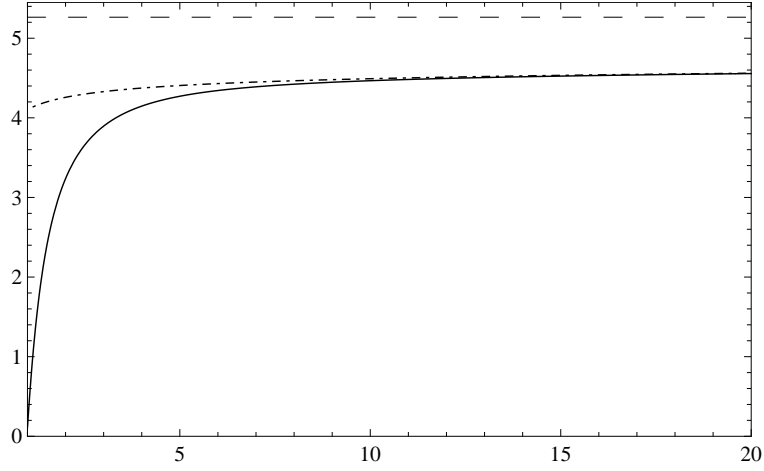}
\caption{The GP pressure (6.43) scaled by $T^4 /3$ is shown as a function of $a=T/T_c \geq 1$ (solid curve).
Its PT part also properly scaled, i.e., $[1 - \alpha_s(T)](SB)$ is shown as dot-dashed curve.
The horizontal dashed line is the general SB constant (A8). It is clearly seen that its shape up to $5T_c$ is determined by
$P_g(T)$, while starting approximately from $10T_c$ the NP contribution from $P_g(T)$ becomes negligible small.}
\label{fig:1}
\end{center}
\end{figure}

We present the GP pressure (6.43) as  well as the lattice pressure
\cite{33} in Fig. 4. Our procedure made it possible to continue lattice data above $3.4T_c$,
as well as to continue them below $T_c$ up to zero temperature. So now
we can predict the value of the GP
pressure, and hence of all other thermodynamic quantities/observables, in the
region of very low temperatures, where lattice uncertainties still
remain very large. One of the interesting features of the lattice simulations \cite{33,36,37} is a rather
slow approach to the common SB
limits (A8) at high temperatures of all the independent
thermodynamic quantities. Within our formalism the regime at high
temperatures is controlled by the running coupling constant
$\alpha_s(T)$ (6.45), which depends on $T$ only logarithmically.
It is instructive to discuss this issue (for further purpose as well)
in more detail. The GP pressure (6.43) above $T_c$ is

\begin{eqnarray}
P(T) &=& \left( 1 -  \alpha_s(T) \right)P_{SB}(T) - 0.55 P_g(T) \nonumber\\
&=& \left( 1 -  \alpha_s(T) \right)P_{SB}(T)
- 0.55 [\Delta^2 T^2 - {6 \over \pi^2} \Delta^2 P'_1 (T) + {16 \over \pi^2} T M(T) + P^s_{PT}(T)], \quad T > T_c,
\end{eqnarray}
where we omit the next-to-leading term $\sim T^{-3}$ in Eq.~(6.46),
since this plays no role in the present discussion. For convenience, in the second line of this equation the gluon pressure $P_g(T)$ (5.2)
is explicitly shown (see discussion below). In dimensionless units (6.19) it looks like

\begin{equation}
{3 P(T) \over T^4}  = [ 1 -  \alpha_s(T)] (SB) - 0.55 {3 P_g(T) \over T^4}, \quad T > T_c,
\end{equation}
because of the relations (A8). The competition between these two terms in Eq.~(7.2) is clearly seen in Fig. 5.
From our numerical results it follows that $P_g(T)$ plays a dominant role in the moderately high temperatures interval up to $5T_c$.
Moreover, $T_c$ is fixed by $P_g(T)$ itself, and the shape of the GP pressure in Fig. 5 just above $T_c$ is determined by it as well.
There is no doubt that the gluon pressure $P_g(T)$ correctly
reproduces the NP structure of the full gluon plasma pressure $P_{GP}(T)$ in the whole temperature range. The addition of the positive PT term (which varies very slowly in the whole temperature range above $T_c$) to $P_g(T)$ in Eq.~(7.2) cannot provide such sharp change in the behavior of the GP pressure just above $T_c$.

On the contrary, in the limit of very high temperatures
the first PT term will become dominant. Indeed, substituting expansion (5.7) into Eq.~(7.2), one obtains

\begin{equation}
{3 P(T) \over T^4} \sim [ 1 -  \alpha_s(T)](SB) - 1.65 \left[ \bar
B_2 \alpha_s \left( {T_c \over T} \right)^2 +  \bar B_3 \left( {T_c
\over T} \right)^3 \right], \quad T \rightarrow \infty,
\end{equation}
where $\bar B_2 = B_2 ( \Delta / T_c)^2$ and $\bar B_3 = B_3 (
\Delta / T_c)^3 + (M / T_c)^3$. Let us remind that the mass gap term $\Delta^2 T^2$
shown in Eq~(7.1) is exactly canceled by the contribution coming from the $M(T)$ composition at high temperatures, and so only
its $\alpha_s$-suppressed counterpart explicitly shown in Eq.~(7.3) survives in this
limit. So in the limit of very
high temperatures the power-type corrections of the second term in Eq.~(7.3) become small starting approximately from $5T_c$.
At approximately $10T_c$ they become simply slighting small in comparison
with the contribution of the first term, which is of a little
(logarithmical) dependence on $T$, see Fig. 5. Just this explains why all the independent thermodynamic quantities
approach their perspective ideal gas limits in the AF way, i.e., slowly and from below, see Fig. 6 and 7 as well.

The exponential suppression of the GP pressure (6.43) in the $T \rightarrow 0$ limit
and its exponential rise close to $T_c$ can be analytically shown from the
expressions (6.43) and (5.5) by putting there $T=T_c - \delta T$
and expanding in powers of a small $\delta = - 1 + (T_c/T)$ in the $T \rightarrow T_c$ limit (for details see section V).
We omit this cumbersome expression, for simplicity, since the exponential rise of GP pressure and other thermodynamic observables
is explicitly seen in Figs. 6 and 7. Concluding, let us underline that above we discussed the main
characteristic features of the pressure: 1). The exponential suppression in the $T \rightarrow 0$ limit. 2). The exponential rise in the
$T \rightarrow T_c$ limit, i.e., its value at  $T_c$. 3). The continuous character across $T_c$.
4). The AF approach to SB constant in the $T \rightarrow \infty$ limit.

\begin{figure}
\begin{center}
\includegraphics[width=10cm]{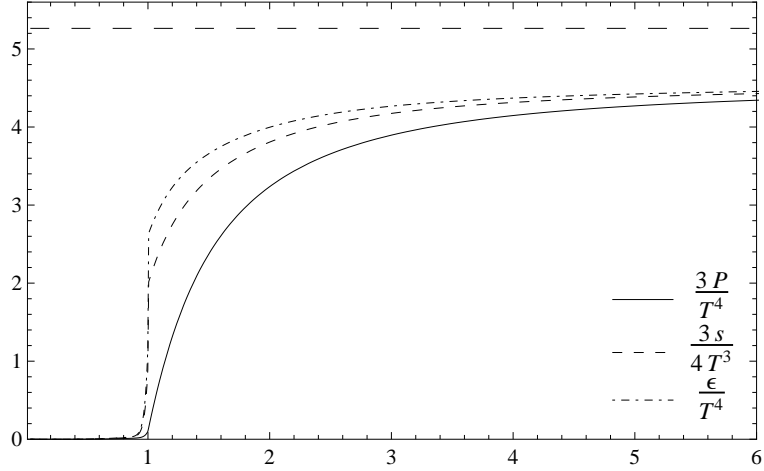}
\caption{The GP pressure (6.43), the entropy and energy densities
(A1) all properly scaled are shown as functions of $T/T_c$. The
finite jumps in densities are clearly seen, and the LH is
$\epsilon_{LH} = 1.41$. Their common SB limit (A8) at high
temperatures (straight dashed line) is rather slowly approaching.}
\label{fig:1}
\end{center}
\end{figure}

\begin{figure}
\begin{center}
\includegraphics[width=10cm]{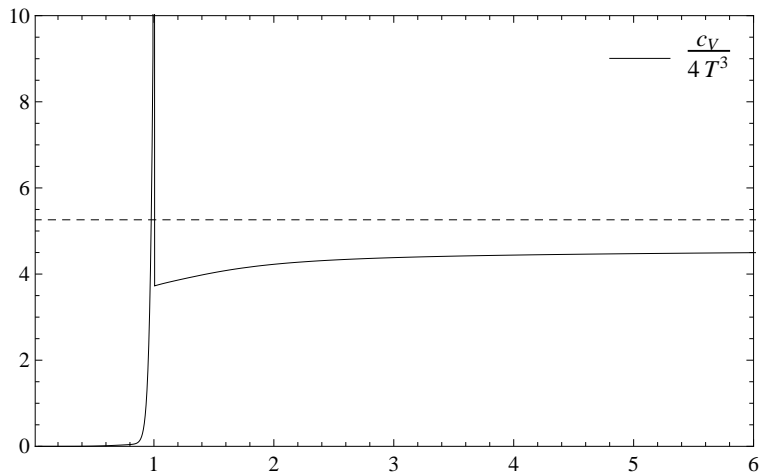}
\caption{The heat capacity (A2) is shown as a function of $T/T_c$.
It has a $\delta$-type singularity (an essential discontinuity) at
$T_c$. It very slowly approaches the general SB constant (A8)
(horizontal dashed line) at high temperatures.  } \label{fig:1}
\end{center}
\end{figure}

\subsection{Energy and entropy densities}

The GP entropy and energy densities (A1) are shown in Fig. 6.
For future purpose, it is instructive to explicitly present the analytical expression for
the energy density valid for high temperatures above $T_c$, namely

\begin{equation}
\epsilon(T) = 3 [1 -\alpha_s(T)]P_{SB}(T) - T \alpha_s'(T) P_{SB}(T) - 0.55 [T P'_g(T) - P_g(T)], \quad T > T_c.
\end{equation}
In derivation of this formula we have used Eq.~(7.1) for the GP pressure as well as some of the relations (A9).
The size of the discontinuity in the energy density, the so-called latent heat (LH) is

\begin{equation}
\epsilon_{LH} = 1.41
\end{equation}
in dimensionless units (see appendices B and E for its definition
and analytical/numerical evaluation, respectively). Let us underline that the same value (7.5) comes from the
independent calculations
of the energy density and the trace anomaly (see below) as it should be, since the pressure itself is a continuous
function across $T_c$, i.e., $ \epsilon_{LH} = \Delta (\epsilon - 3 P) / T_c^4 = \Delta \epsilon / T_c^4$
(here $\Delta$ is not mass gap, see appendix B, as well as discussion in subsection E).
This means that the first-order phase transition in the GP is analytically confirmed
for the first time, in complete agreement with thermal $SU(3)$ YM
lattice simulations \cite{33,36,37,42} (and references
therein). The reason of such sharp changes at $T_c$ in the
derivatives of the GP pressure is that its exponential rise below
$T_c$ is changing to the polynomial rise above $T_c$ in order to
reach finally the SB limit. The value (7.5) is in fair agreement
with lattice ones in \cite{9,33,37,43,44,45} (and references therein).
This agreement is not a trivial thing, since, we have adjusted our
analytical numerical simulations with those of lattice ones in
\cite{33} only for the pressure. First of all, the energy and
entropy densities (being derivatives of the pressure), nevertheless, are an independent thermodynamic observables.
Secondly, the lattice results heavily depend on how the continuum
limit is to be taken and on other details of the above-cited lattice
simulations. For example, the lattice data points closest to $T_c$ for the entropy density may still be affected by an
upward finite-volume effect \cite{9}, while the pressure is a continuous function across $T_c$, as underlined above.
The slow approach of the energy and
entropy densities to their common SB limit (Fig. 6) has already been
explained in the previous subsection.

\subsection{Heat capacity}

The last independent thermodynamic quantity the heat capacity,
defined in Eq.~(A2), is shown in Fig. 7. It is always a smoothly
growing function of $T$, both below and above $T_c$, while at $T_c$
it has a $\delta$-type singularity (an essential discontinuity) due
to the expression (B5). It very slowly approaches the common SB
limit (A8) at high temperatures.

\subsection{Conformality, conformity and the velocity of sound squared}

The GP pressure versus the GP energy density, i.e.,  $P(\epsilon)$,
is present in Fig. 8. The size of the LH and a rather rapid approach
to conformality are clearly seen. We distinguish between
conformality here and conformity defined in Eq.~(A4), though
numerically in the limit of high temperatures they are the same.
Conformity itself is shown in Fig. 9. It has a finite negative jump
at $T_c$ because of a jump in the energy density at this point, and
it rather rapidly approaches its SB limit (A9) at high temperatures.
However, its most interesting feature is a non-trivial dependence on
$T$ below $T_c$, which has been fixed explicitly in $SU(3)$ GP for
the first time. Its shape can be due to the fact that
conformity is the ratio of the independent thermodynamic quantities
(for it the exponential suppression at low temperature is not
mandatory). If the existence of the protuberance in the region
$\sim (0.2-0.4)T_c$ is a physical effect, then it has to be consequence of the complicated NP structure of the gluon
pressure $P_g(T)$ and its derivatives. It dominates the structure of
the GP pressure below $T_c$. If it is a mathematical artefact, then it is due to the SB-type terms.
Their penetration so deeply into the low-temperature region is very small (see Eqs.~(6.43)-(6.46)), indeed.
In principle, the shape of the curves below $T_c$ (see Figs. 9 and 10) may be
changed (or not?) by use of more terms in the summation over $i$ in the relations (6.5).
In any case, the corresponding numbers will be rather small, and the problem
(the existence of the protuberance) seems not to be so important from the numerical point of view (at least, beyond
the accuracy of our numerical simulations, as discussed above). The velocity of sound squared (A3) is shown
in Fig. 10. Below $T_c$ it behaves very similarly to conformity (Fig.
9), since the latter one mimics its properties. The principal
difference from conformity is that at $T_c$ it is zero because of
the heat capacity having the above-mentioned $\delta$-type
singularity at this point. It rather rapidly approaches its SB limit
(A9) at high temperatures.

\begin{figure}
\begin{center}
\includegraphics[width=10cm]{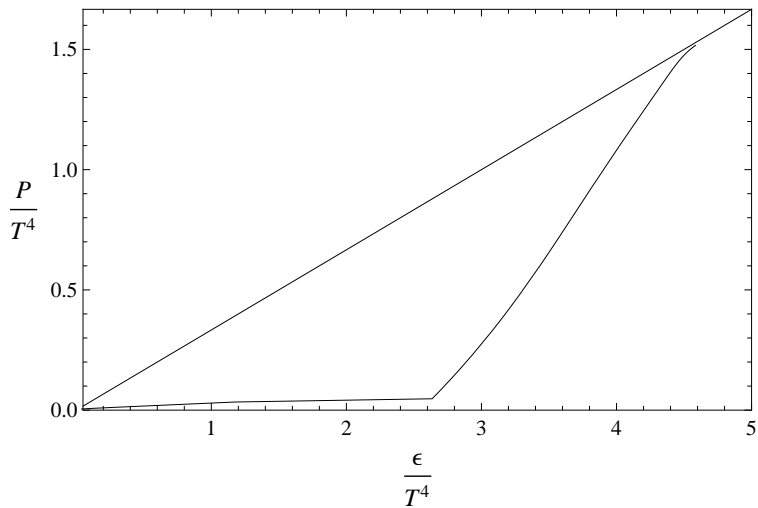}
\caption{$P(\epsilon)$ EoS (solid line) and rather rapid approach to
$conformality = 1/3$ (diagonal thin line).} \label{fig:1}
\end{center}
\end{figure}

\begin{figure}
\begin{center}
\includegraphics[width=10cm]{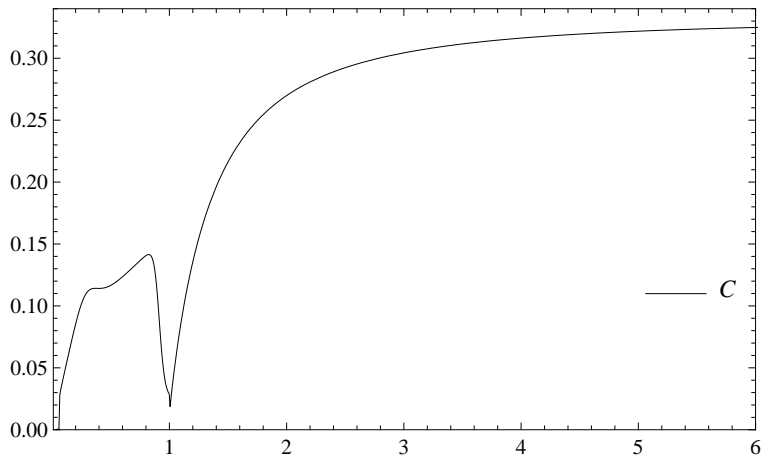}
\caption{Conformity (A4) is shown as a function of $T/T_c$. It is
zero at $T=0$, and has a finite negative jump at $T_c$ due to a jump
in the energy density at this point. It shows a non-trivial
dependence on $T$ below $T_c$, and rather rapidly approaches the SB
limit (A9) at high temperatures.} \label{fig:1}
\end{center}
\end{figure}

\begin{figure}
\begin{center}
\includegraphics[width=10cm]{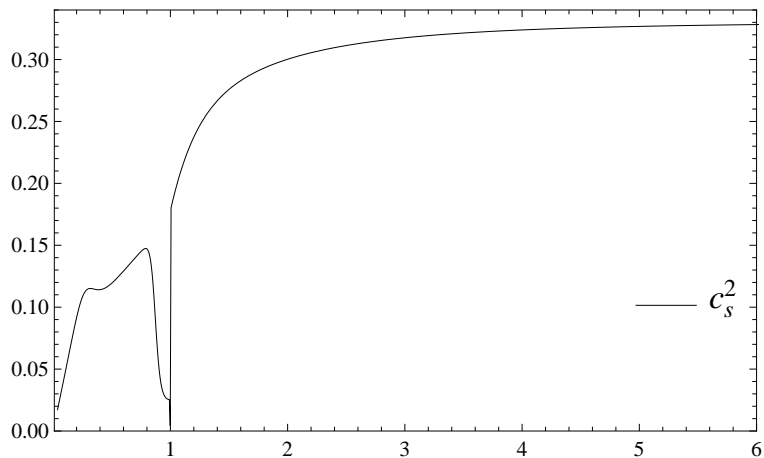}
\caption{The velocity of sound squared (A3) is shown as a function of $T/T_c$.
It shows a non-trivial dependence on $T$ below $T_c$, while at $T=0$ and at $T=T_c$ it is zero.
It rather rapidly approaches the SB limit (A9) at high temperatures. }
\label{fig:1}
\end{center}
\end{figure}

\subsection{Trace anomaly relation and the gluon condensate}

The trace anomaly, defined in Eq.~(A5), is especially sensitive to
the NP effects, since the corresponding pure PT contributions are exactly cancelled in this composition,
as it follows from the relations (7.1) and (7.4) (but not their derivatives, see below).
Properly scaled it is shown in Fig. 11.
The rapid rise of the peak (due to the LH in the energy density, see
Fig. 6) is exactly placed at $T_c$, and it is about $2.5$. In all
lattice calculations it peaks at about $1.1T_c!$
\cite{33,34,36,37,44}, and it is about $2.6$, and almost coincides with our value in \cite{37}. The wrong position of the
lattice trace anomaly peak can be due to an ultraviolet cutoff, the finite volume effects, etc.
In this connection let us indeed remind that in lattice simulations at any
temperature it is necessary to finally go to the continuum
(physical) limit, namely lattice spacing goes to zero and then the infinite volume limit
should be taken. These are nothing else but the removal of the ultraviolet and infrared cutoffs which is the
part of the renormalization procedure \cite{46,47}.
It seems to us that our analytical method resolves this $SU(3)$ lattice thermodynamics artefact.

Just above $T_c$ and up to rather high temperatures $(4-5)T_c$ the NP effects due to the mass gap are
still important in the trace anomaly. Fig. 11 demonstrates rather complicated dependence of the
trace anomaly on the mass gap and the temperature in this interval.
Indeed, the trace anomaly equation (A5), on account of the expressions (7.1) and (7.4), and divided by $T^4$ is

\begin{equation}
{I(T) \over T^4}= {\epsilon(T) - 3 P(T) \over T^4}   = - {1 \over
3}T \alpha_s'(T) (SB) - 0.55 {[T P'_g(T) - 4 P_g(T)] \over T^4}, \quad T > T_c,
\end{equation}
where the derivative of the PT effective charge $T \alpha_s'(T)$ is given in the relation (D6).
In Fig. 11 the trace anomaly relation with the derivative of the pure PT contribution being subtracted is shown as a dashed line
(it is defined as follows: $I^s(T) / T^4 = I(T) / T^4 + (1 /3) T \alpha_s'(T)(SB)$).
This means that the main contribution to the trace anomaly comes from the second NP term in Eq.~(7.6),
and it is not a simple power-type fall off.
It is mainly due to the complicated dependence of the gluon pressure $P_g(T)$
on the mass gap and the temperature in this region, where it cannot be
approximated by some simple power-type expression. However, this is possible to do in the limit of very high
temperatures approximately above $(4-5)T_c$. Substituting the asymptotic (5.7) and its derivative
into the previous equation and doing some algebra, one obtains

\begin{equation}
{I(T) \over T^4}  \sim - {1 \over 3} T \alpha_s'(T) (SB) +  1.1 \bar B_2 \alpha_s \left( {T_c \over T} \right)^2 +
1.65 \bar B_3 \left( {T_c \over T} \right)^3, \quad T \rightarrow \infty,
\end{equation}
where $\bar B_2 = B_2 ( \Delta / T_c)^2$ and $\bar B_3 = B_3 (
\Delta / T_c)^3 + (M / T_c)^3$, and for the coefficients $B_2, B_3$
and the quantity $M$ see text at the end of section V.

Let us now discuss one important problem in connection with the trace anomaly. The jump or, equivalently, the latent heat calculated through the
energy density
and the trace anomaly, should be the same (7.5) due to continuous character of the GP pressure across $T_c$. However, for the trace anomaly defined
by the subtraction of all types of the PT contributions (the derivative of the PT effective charge in Eq.~(7.4)) it is less than the value (7.5).
This is clearly seen in Fig. 11. Calculated in appendix E, numerically it is
$\epsilon_{LH}^s = \epsilon_{LH} - 0.0899 \times 0.95366 = 1.41 - 0.0899 \times 0.95366 = 1.324266$.
If the subtraction of all types of the PT contributions in the trace anomaly relation at zero and non-zero temperature
can be justified \cite{7} (and references therein), there are no such arguments to do the same in the energy density itself.
We are going to discuss this problem in more detail in a separate investigation.
In the forthcoming paper we also
intend to investigate the trace anomaly scaled by $T^2T^2_c$, i.e., $I(T)/T^2T^2_c$ \cite{31}, as well as $I(T)/T^4$
as a function of $(T_c/T)^2$, and discuss them following paper \cite{34}. We will do this by applying our approach to analytically describe results of
the precision $SU(3)$ lattice thermodynamics for a large temperature range in \cite{37}.

In close connection with the trace anomaly is the gluon condensate
defined in Eq.~(A6) and shown in Fig. 12. It approaches zero from
below, so it gains very small negative values at high temperatures
(fixed also by lattice simulations in \cite{36}). This is due to the
fact that the trace anomaly enters Eq.~(A6) with negative sign.

\begin{figure}
\begin{center}
\includegraphics[width=10cm]{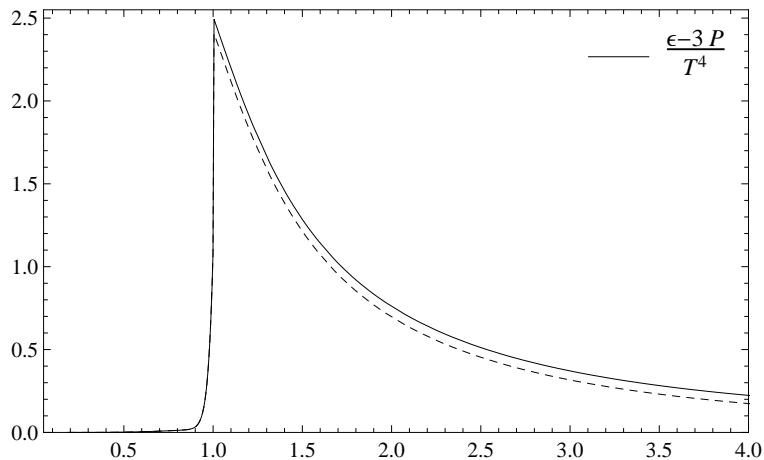}
\caption{The trace anomaly (A5) properly scaled is shown as a function of $T/T_c$. Its subtracted counterpart is shown as dashed line (see discussion around Eq.~(7.6))}
\label{fig:1}
\end{center}
\end{figure}

\begin{figure}
\begin{center}
\includegraphics[width=10cm]{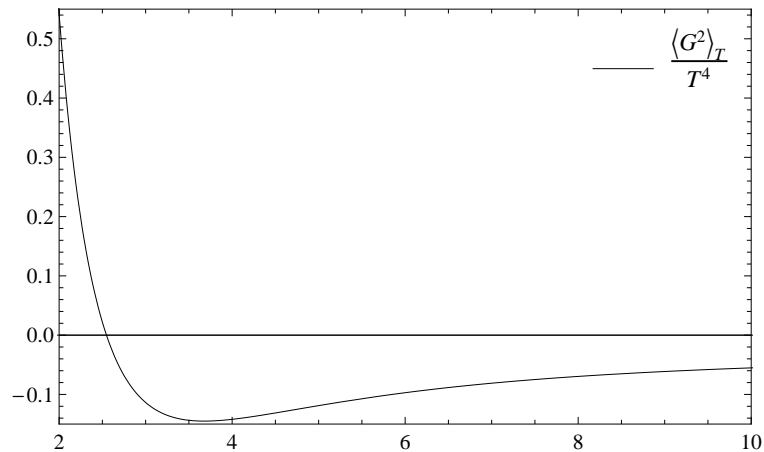}
\caption{The properly scaled gluon condensate at non-zero temperature (A6)
is shown as a function of $T/T_c$. It approaches zero from below at high temperatures.}
\label{fig:1}
\end{center}
\end{figure}

\section{The dynamical structure of $SU(3)$ GP}

The GP pressure (6.43) describes the three different types
of massive gluonic excitations. They are: $\omega'$ and $\bar
\omega$ with the effective masses $m'_{eff}= 1.17 \ \GeV$ and $m_{eff} = 0. 585 \ \GeV$, respectively. The third
one is again $\bar \omega$ which propagation, however, is suppressed by the
$\alpha_s$-order. We have denoted it as $\alpha_s \cdot \bar \omega$. We can treat it as a new massive
excitation, but with the same effective mass $\bar m_{eff} = 0. 585 \ \GeV$. Both effective masses are due to the mass gap $\Delta^2$,
i.e., they have not been introduced by hand. The mass gap itself is
dynamically generated by the nonlinear interaction of massless gluon
modes \cite{7}. The first effective mass $m'_{eff}$ is comparable
with the masses of scalar glueballs \cite{48,49}. The effective mass
$\bar m_{eff}$ might be identified with an effective gluon mass of
about $(500-800) \ \MeV$, which arises in different quasi-particle
models \cite{48} (and references therein). We also have the two
different massless gluonic excitations $\omega$, conventionally
denoted as $\omega_1$ and $\omega_2$. The former describes the
massless gluons, propagating in accordance with the integral
$P_1(T)$ in Eq. (3.2). The latter one describes the massless gluons,
propagating in accordance with the integral $P_2(T)$ in Eq. (3.3).
Let us remind that the integral (3.2) should be multiplied by $(6
/\pi^2) \Delta^2$, and all other integrals (3.3) are to be
multiplied by $(16 / \pi^2) T$, when one speaks about different NP
contributions to the pressure (3.1). The  propagation of the massive
gluonic excitation $\alpha_s \cdot \bar \omega$ has to be understood
in the same way (i.e., the corresponding integral should be
multiplied by overall numerical factor, see Eq.~(5.4)).

In the low-temperature $T \rightarrow 0$ limit all these excitations
are exponentially suppressed, since the low-temperature asymptotic
of the GP pressure is determined by the gluon pressure $P_g(T)$, see
Eq.~(5.5). The dependence on the mass gap and the
temperature of the terms, describing their propagation at finite
temperatures, is rather complicated. In the high-temperature $T
\rightarrow \infty$ limit the suppression is a power-type, if the
GP pressure is scaled by $T^4 /3$. Hence in this limit the PT
excitations play dominant role, see discussion in section VII. Let us remind that in
both limits a non-analytical dependence on the mass gap occurs, but
it is not an expansion parameter like $\alpha_s$.

It is important to understand that the above-mentioned effective
masses are not the pole masses which appear in the corresponding
propagators (see, for example \cite{22,23,50}). This means that we cannot
assign to the corresponding massive excitations a meaning of being
physical particles. They have to be treated rather as
quasi-particles (see discussion below). They appear through the
corresponding gluon mean numbers, something like the quark chemical
potentials. Indeed, from Eq.~(3.6) one gets

\begin{equation}
N'_g \equiv N_g(\beta, \omega') = {1 \over e^{\beta\omega'} -1} =
{1 \over e^{\beta\sqrt{\omega^2 + m'^2_{eff}}} -1} =  {1 \over e^{\beta(\omega - \mu'_g)} -1}
\end{equation}

where we introduce the fictitious gluon "chemical potential"
$\mu'_g$. It has to satisfy the following equation $\mu'^2_g - 2
\omega \mu'_g - m'^2_{eff} =0$, which has the two independent
solutions: $\mu'_g = \omega \pm \omega' = \omega \pm \sqrt{\omega^2
+ m'^2_{eff}}$, leading, nevertheless, to the same effective mass
$m'^2_{eff}$, but only the solution $\mu'_g = \omega - \omega'$ is
compatible with Eq.~(8.1). By making the replacement $\omega'
\rightarrow \bar \omega$ in Eq.~(8.1), we can treat the massive
gluonic excitation $\bar \omega$ in the same way as $\omega'$.
Again, one has the two independent solutions for the fictitious
gluon "chemical potential" $\bar \mu_g = \omega \pm \bar \omega =
\omega \pm \sqrt{\omega^2 + \bar m^2_{eff}}$, leading, nevertheless,
to the same effective mass $\bar m^2_{eff}$, but only the solution
$\bar \mu_g = \omega - \bar \omega$ will be compatible with the
corresponding Eq.~(8.1). In the excitation $\alpha_s \cdot \bar
\omega$ the effective mass $\bar m^2_{eff}$ appears not only through
the corresponding gluon mean number, but in a more complicated way,
see Eq.~(5.4). For convenience, we denote its "chemical potential"
as $\alpha_s \cdot \bar \mu_g$. All three gluon "chemical
potentials" $\mu'_g$, $\bar \mu_g$ and  $\alpha_s \cdot \bar \mu_g$
differ from each other by the corresponding effective masses and by
the ranges for $\omega$ (see integrals $P_3(T)$, $P_4(T)$ in
Eq.~(3.3) and integral in Eq.~(5.4)). The corresponding
gluon "chemical potentials" for the two massless excitations
$\omega_1$ and $\omega_2$ are zero, i.e., $\mu_1=\mu_2=0$ with the
same range for $\omega$, see integral (3.2) and the first of
integrals (3.3).

In principle, we can interpret our effective excitations as the
gluon "flavors", but better to use the term  "species". So we have
the five different gluonic species, which are present in the GP.
Contrary to the quark flavors, all our species are of NP dynamical
origin, since in the PT $\Delta^2=0$ limit they disappear from the
GP spectrum (the dependence of the massive species $\mu'_g$, $\bar
\mu_g$ and  $\alpha_s \cdot \bar \mu_g$ on the range for $\omega$
only confirms this). In other words, it is better to treat our
massive excitations/species as some kind of quasi-particles, created
by the self-interaction of massless gluon modes at non-zero
temperature, i.e., consisting of the GM only. That these masses are
very close to scalar glueballs and Debye screening masses may or may
not be a coincidence, but there are no any other massive excitations
in the GP from the very beginning. Let us remind that their values are in
good agreement with established thermal PT QCD results \cite{22,48,50}.

At present, nobody can definitely
answer the question why some gluons acquire a mass and some others
not. At finite temperatures some gluon fields may intensively
interact with each other, leading thus to the formation of stable
gluon field configurations, the so-called "stationary" states with the minimum of energy.
A possible existence of such kind of states of purely transversal
virtual gluon field configurations in the QCD vacuum at zero temperature has been discussed
in detail in \cite{7}. At non-zero temperature the above described stationary states
might be also formed/created, and effectively they can be considered as
the massive gluonic excitations. If the self-interaction of massless
gluon modes is very intense (possibly creating something like bound-states of gluons),
and an effective mass is big enough then
such a stable configuration can be treated as a "glueball". If the
self-interaction is not intense then an effective mass is not so
big. Such a configuration may be considered as a "massive" gluon. If
the self-interaction can be neglected, then the gluons remain
massless. In any case, the different types of the massive and
massless gluonic excitations of the dynamical origin will necessary
appear at non-zero temperature \cite{7,48} (and references therein).

The GP pressure (6.43) assumes the presence of the two different PT
massless excitations, propagating above $T_c$. The first one is described by the SB term $P_{SB}(T)$
itself, while the second one by its AF correction denoted as $\alpha_s(T) \cdot P_{SB}(T)$.
It is necessary to point out that the AF correction has been empirically restored to the GP
pressure, and it was agreed with lattice results in a very short temperature interval in \cite{7}.
At the same time, in the present investigation it is, in fact, the numerical solution of the renormalization
group equation of motion (see appendix D). And the behavior of the GP pressure is agreed
with all the possible lattice results above $T_c$, as described in subsection VI.B.
We also necessarily have the two types of the exponentially suppressed terms,
contributing to the GP pressure below $T_c$, conventionally called the SB-type terms.
In our opinion they have a little physical sense, more mathematical one (due to the above discussed normalization condition
of the free PT vacuum to zero). They were needed to ensure the continuous character of the GP pressure below and across $T_c$.
These are $P_g(T)$ and $P_{SB}(T)$ with its AF correction which are responsible for the NP and non-trivial PT
dynamics, respectively, in the GP pressure (6.43).

$P_g(T)$ describes the changes in
the regime of the GP pressure's behavior near $T_c$, namely the exponential
rise transforms to the polynomial one, providing a continuous
transition of the GP pressure across $T_c$ with the help of the SB-type terms, as underlined just above.
All the NP massive and massless gluonic excitations/species have not been introduced by hand; on the
contrary, they are of the dynamical origin due to the confining
effective charge (see appendix C). They are described and accounted for by the gluon pressure $P_g(T)$.

The exponential rise of all the independent thermodynamic quantities
in the transition region $\sim (0.8-1)T_c$ clearly seen in Figs. 6
and 7 indicates that near $T_c$ a dramatic increase in the number
of effective gluonic degrees of freedom will appear (and we know
that this is so indeed, see remarks in section V and at the end of subsection VII.A).
The massive excitations/species will begin to rapidly dissolve. This will lead to drastic changes in
the structure of the GP. A change in this number is enough to
generate pressure gradients, but not enough to affect the pressure
itself. It varies slowly and therefore remains continuous in this
region. At the same time, the pressure gradients such as the energy
and entropy densities, etc., undergo sharp changes in their
behavior, having different types of discontinuities at $T_c = 266.5 \ \MeV$. Thus
$SU(3)$ GP has a first-order phase transition with latent heat
$\epsilon_{LH} = 1.41$. Of course, not all the
massive excitations will be dissolved in the transition region. Some
of them will remain above $T_c$ together with other gluonic
excitations and effective gluonic degrees of freedom, which may
include the above-mentioned different PT contributions as well as
gluon condensates. This forms a mixed phase around $T_c$
\cite{51}. One can conclude that the NP physics of the mixed phase
(the temperature interval approximately $(0.8 \rightarrow (4-5))T_c$) is well now understood. The
region of low temperatures, where all the independent thermodynamic
variables are exponentially suppressed at $T \rightarrow 0$, is also now
under control. In the mixed phase the $SU(3)$ GP can be considered as
being in the strong coupling regime, so that its behavior in this region is different
from the behavior of a gas of free massless gluons. Beyond it the NP effects become
small, and the GP can be considered as being in the weak coupling
regime. However, all the independent thermodynamic quantities approach rather
slowly their respective SB limits at high temperatures, Figs. 6 and 7.
The thermodynamic quantities which are the ratios of the
corresponding independent counterparts rather rapidly approach their
respective SB limits at high temperatures, Figs. 8,9 and 10.
The structure of the GP will be mainly determined by the SB
relations (A8)-(A10) between all the thermodynamic quantities
at very high temperatures only.

\section{Conclusions}

A general approach how to analytically describe and understand $SU(3)$ lattice
thermodynamics in the whole temperature range $[0, \infty)$  is formulated and used.
It is based on the effective potential approach for composite operators properly generalized
to finite temperature. This makes it possible to introduce into this formalism a dependence
on the mass gap \cite{15,52}, which is responsible for the large-scale dynamical structure of the QCD ground state \cite{7}.
The gluon pressure $P_g(T)$ (5.2) as a function of the mass gap $\Delta^2$, analytically derived and
numerically calculated within this approach in our previous works \cite{7,16}, has been briefly described in sections II, III, IV and V.
It fixes the value of the characteristic temperature $T_c= 266.5 \ \MeV$.

In section VI we explain why and how EoS for the gluon pressure $P_g(T)$ has to be changed to the GP pressure $P_{GP}(T)$
in the most general way. The gluon pressure $P_g(T)$ is a necessary analytical and dynamical input information for the GP pressure (6.1).
On the other way around, the lattice pressure \cite{33} is a main numerical input information to use in order to fix
the functions $L(T)$ and $H(T)$ in Eq.~(6.1).

In subsection VI.A we proposed and developed a method of analytical simulations
which allows one to express such introduced functions $L(T)$ and $H(T)$ in terms of basic functions $P_g(T)$ and $P_{SB}(T)$, multiplied by
the so-called simulating functions $\phi_l(T), \ \phi_h(T)$  and $f_l(T), \ f_h(T)$, respectively, see relations (6.3).
They are to be necessary represented by the corresponding asymptotics
of the gluon mean number (3.6) in the low- and high temperature limits. This makes it possible to reproduce lattice data in any requested
temperature interval and to ensure the correct SB limit for all the thermodynamic obsrvables/quantities as well.

In subsection VI.B we have performed the numerical simulation of the GP pressure (6.17) above $T_c$ with the help of Eq.~(6.30)
in order to fix the function $\phi_h(T)$ (6.35) in accordance with the lattice pressure \cite{33} in this region by using the LMS method.
Our procedure, described in detail in this subsection, makes it also possible to continue the lattice pressure to the region
of very high temperatures with the help of Eq.~(6.36) see Fig. 2.

In subsection VI.C we have performed the numerical simulation of the GP pressure (6.37) below $T_c$ with the help of Eq.~(6.40)
in order to fix the fitting parameters $\mu_1, \ \mu_2$ and $A_2$ in accordance with lattice data \cite{33} in this region,
but only very close to $T_c$. Our procedure makes it also possible to continue the calculation of the GP pressure to very low temperatures
with the help of Eq.~(6.42), where convincing lattice data does not exists at all. Thus, we can predict the behavior of the
lattice pressure curve up to zero temperature, knowing only its behavior very close to $T_c$, see Fig. 3.

In subsections VI.D and VII.A the analytical expression (6.43) reproducing the lattice pressure
\cite{33} in the whole temperature range $[0, \infty)$ as a function of the mass gap $\Delta^2$ is present and discussed, see Fig. 4.
According to such obtained analytical expression, the corresponding lattice pressure is exponentially suppressed at low temperatures,
smoothly approaching zero in the $T \rightarrow 0$ limit, i.e., having no finite zeroes below $T_c$. It shows exponential rise close to $T_c$, while being
continuous across $T_c$ and approaches its SB limit at high temperatures in AF way. Thus, in general, it satisfies all the established thermodynamics limits.
In other words, the GP pressure (6.43) is, in fact, the lattice pressure \cite{33} analytically expressed as a function of
the mass gap and temperature and properly continued to the regions of very low and high temperatures. Let us also emphasize that Eq.~(6.43)
is a unique solution as well, since all other different combinations of numbers $n$ and $m$ in the relations (6.5), which finally determine the numerical
structure of the GP pressure below $T_c$ explicitly shown in Eq.~(6.43), failed to reproduce all the necessary requirements described above. At the same time, the functions $f_h(T)$  and $\phi_h(T)$ in the relations (6.3), which determine the numerical
structure of the GP pressure above $T_c$, have been uniquely fixed as described in subsections VI.A and VI.B, respectively.

In section VII using further all the thermodynamic relations shown in appendixes A and B, we were able to calculate all other thermodynamic observables
as functions of the mass gap. This makes it possible to analytically investigate $SU(3)$ lattice thermodynamics in the whole temperature
range on a general dynamical ground by using only three independent fitting parameters, mentioned above.
The parameters (6.34) were not called as fitting ones, since their numerical values have been uniquely fixed by LMS method.

In section VIII we describe the dynamical structure of the GP which emerges within our approach. A few points are necessary to
underline. The massive excitations which appear in our picture are not pole masses. The main dynamical source of these effective masses
is the self-interaction of massless gluon modes \cite{7}. Expressed in terms of the mass gap they are to be treated as quasi-particles, indeed,
since they appear through the corresponding gluon mean numbers. A few remarks as a subject for the discussion are present
in order to answer the question why some gluons acquire a mass and some others not. We also give the explanation why the GP should undergo
a first-order phase transition in our picture. At long last, it is due to principally different asymptotics of the gluon pressure $P_g(T)$ in the low- and
high temperatures limits; exponential and power types, respectively (see section V). If for the GP pressure itself this difference plays no role, for its derivatives it becomes important, leading to the discontinuities of different types (see section VII and appendixes B and E).
There are no doubts that $P_g(T)$ correctly describes not only the dynamical context of the lattice pressure (6.43)
but its analytic structure as well. It also correctly describes its low-temperature asymptotic properties, and makes it possible to restore
its SB limit at high temperature in a self-consistent way.

So we know now what is the physics behind all the lattice curves and their numbers within our
approach. It is a general one, indeed, since knowing the pressure, any other thermodynamic quantity can be calculated from it.
It is worth emphasizing once more here that Eq.~(6.43) is nothing else but the analytical version of the lattice pressure \cite{33}.
Any lattice thermodynamic quantity, calculated in any given temperature interval can be analytically expressed as a function
of the mass gap within our approach, which is evidence of its flexibility, in our opinion.
We have explicitly shown how lattice and analytical simulations have to be united in
order to describe and understand the lattice thermodynamics on the general dynamical ground (the mass gap) and
in the whole temperature range $[0, \infty)$. To our best knowledge such kind of investigation has been done for the first time.

In this connection a few remarks are in order. Analytical formalism updated and further developed in the present paper has been formulated first in
our book \cite{7}. There it has been applied in order to cover rather short temperature interval $a=2.8-3.4$ which included only 33 lattice data points. As a result, the LMS method gave the value of the GP pressure at $T_c$ approximately two times bigger than the actual lattice pressure is at $T_c$ in \cite{33}. At the same time, here we were able to cover all the possible (above $T_c$) temperature interval
$a=1-3.4$ which included 162 lattice data points, increasing thus drastically the accuracy of the LMS method.
Also, we have correctly reproduced the lattice pressure values below $T_c$ but
close to $T_c$ and at $T_c$ itself, which we failed to do
in \cite{7}, as mentioned above. All this led to rather different pictures obtained previously and here
(compare Figs. 9.3.1, 9.3.3 and 9.C.1 in \cite{7} with Figs. 2,3, and 15 in this paper). Moreover, in \cite{7} the approach to SB limit has been
fixed empirically, while in this paper it has been fixed by the numerical, in fact, solution of the renormalization group equation for the
PT effective charge (see appendix D). Despite the similarity of some figures there and here,
the numbers behind them are significantly different (compare Eqs.~(9.3.39-9.3.41) in \cite{7}
and Eqs.~(6.43-6.46) here). In addition to this, let us note that Table I created here has not been created in \cite{7}.
Thus, the interpretation of the present results as analytical description of all the lattice thermodynamics
in the whole temperature range and on general dynamical ground (mass gap) is correct here, while in \cite{7} we could
not make such a clime, and we did not it.
In fact, the previous description \cite{7} is a particular case of the present investigation, since the first one coincides with the second one
only in rather short temperature interval, mentioned above.
This is a principal distinction between the previous model-building and present general approach descriptions, and hence
the interpretation of the present results is completely different from those in \cite{7}, indeed.
All this justifies our general statements made just above.

The next step will be to analytically investigate
within this approach the $SU(3)$ lattice pressure calculated in \cite{37}.
It has to be done in a separate article due to some special aspects of lattice calculations in the above-mentioned paper (see appendix D as well).
Completing this program and drawing some general conclusions from this and forthcoming papers, we will be able to compare our general approach
with others \cite{8,9,10,11,12,53,54,55} (and references therein).

The analytic formula for the lattice pressure, and hence of any other thermodynamic
quantity, will drastically simplify the investigation and solution of the relativistic hydrodynamics equations of motion
\cite{21,56,57,58,59,60} in the case of the pure GP. This will allow to conclude whether it is a perfect fluid or not.
This work is also our future aim.

We are also planning to extrapolate this approach to the quark degrees of freedom in order
to analytically describe and understand already existing \cite{1,2,3,4,5,6,17,61,62,63,64} lattice QGP EoS
and future finite density quantum field theories \cite{65} (and references therein).
Let us emphasize that we can do this only after putting YM thermodynamics on a firm physical (analytic) and numerical (lattice)
joint grounds.

\begin{acknowledgments}

We especially thank M. Panero for providing us with the lattice data
from his paper \cite{33}. We would like to thank J. Rafelski for useful correspondence and discussion.
We thank R. Pisarski and A.S. Kronfeld for bringing our attention to papers \cite{30} and \cite{47},
respectively. Our thanks also go to T. Bir\'o, T. Csorg\"o, P.
V\'an, G. Barnaf\"oldi, A. Luk\'acs, M. Nagy, J. Nyiri, S. Pochybova
for useful discussions, remarks and help. V.G. and A.S. are grateful to V. Kiguradze,
N. Partsvania for constant support and interest.
V.G. and A.S. acknowledge the support by the Hungarian National Fund (OTKA) 77816 and 31520 (P. L\'evai).
Partial support comes from "NewCompStar", COST Action MP1304.
M.V. was also supported by the J\'{a}nos Bolyai Research Scholarship of the Hungarian
Academy of Sciences.

\end{acknowledgments}

\appendix

\section{Main thermodynamic quantities}

Together with the pressure $P(T)$, the main thermodynamic
quantities are the entropy density $s(T)$ and the energy density
$\epsilon(T)$. The general formulae which connect them are
\cite{22}

\begin{eqnarray}
s(T) &=& {\partial P(T) \over \partial T}, \nonumber\\
\epsilon(T) &=&  T \left( \partial P(T) \over \partial T \right) -
P(T)= T s(T) - P(T)
\end{eqnarray}
for pure YM fields, i.e., when the chemical potential is equal
to zero. Let us note that in quantum statistics the pressure
$P(T)$ is nothing but the thermodynamic potential $\Omega(T)$
apart from the sign, i.e., $P(T) = - \Omega(T) > 0$.

Other thermodynamic quantities of interest are the heat capacity
$c_V(T)$ and the velocity of sound squared $c^2_s(T)$, which are defined
as follows:

\begin{equation}
c_V(T) = { \partial \epsilon(T) \over \partial T}  = T \left(
\partial s(T) \over \partial T \right),
\end{equation}
and

\begin{equation}
c_s^2(T) = { \partial P(T) \over \partial \epsilon(T)}  = { s(T)
\over c_V(T)},
\end{equation}
i.e., they are defined through the second derivative of the pressure. The conformity

\begin{equation}
C(T) = { P(T) \over \epsilon (T)}
\end{equation}
mimics the behavior of the velocity of sound squared (A3) but
without involving such differentiation.

A thermodynamic quantity of special interest is the thermal
expectation value of the trace of the energy momentum tensor. This
trace anomaly relation measures the deviation of the difference

\begin{equation}
I(T)= \epsilon(T) - 3P(T) = T^5  {\partial \over \partial T} \left( { P(T) \over T^4} \right)
\end{equation}
from zero at finite temperatures, while in the high temperature limit it must vanish
according to the SB relations (see below). As a consequence
it is very sensitive to the NP contributions to the EoS.
It is also known as the interaction measure and denoted as in Eq.~(A5). We use both notations since they are equivalent.
Let us note that in close connection with this thermodynamic identity
is the other one, namely

\[
T {\partial \over \partial T} \left( { s(T) \over T^3} \right)  = { 1 \over T^3}  {\partial \over \partial T} \left( \epsilon(T) - 3 P(T) \right),
\]
which can be easily verified.

The trace anomaly relation (A5) assists in the temperature dependence of the gluon condensate \cite{36,66}

\begin{equation}
<G^2>_T = <G^2>_0 - [ \epsilon(T) - 3P(T)],
\end{equation}
where $<G^2>_0 \equiv <G^2>_{T=0} =\langle{0} | (1 / 4)G^a_{\mu\nu}
G^a_{\mu\nu} | {0}\rangle  = 0.1052 \ \GeV^4$ denotes the gluon
condensate at zero temperature. Its numerical value is discussed in \cite{7}.

The so-called enthalpy density \cite{67} is defined as follows:

\begin{equation}
e(T) = T {\partial P(T) \over \partial T} = T s(T) = \epsilon(T) + P(T).
\end{equation}
This sum is of interest and importance, since it appears in the above-mentioned
relativistic hydrodynamics equations of motion, making them highly non-linear ones. The curve
for it is shown in Fig. 6, since from the definition (A8) it follows
that $3e(T) / 4 T^4 = 3 T s(T) / 4T^4= 3 s(T) /4 T^3$.

\hspace{2mm}

{\bf The general SB constant/limit}

\hspace{2mm}

The high-temperature behavior of all the thermodynamic quantities
is governed by the SB ideal gas limit, when the
matter can be described in terms of non-interacting massless
particles (gluons). In this limit these quantities satisfy the
special relations, namely

\begin{equation}
{3P_{SB}(T) \over T^4} = {\epsilon_{SB}(T) \over T^4} =
{3 s_{SB}(T) \over 4 T^3} = {c_{V(SB)}(T) \over 4 T^3} = (SB) = { 24 \over
45} \pi^2 \approx 5.2638, \quad T \rightarrow \infty,
\end{equation}
and

\begin{equation}
T P'_{SB}(T) = 4 P_{SB}(T)= {4 \over 3} \epsilon_{SB}(T) = T s_{SB}(T) = { 1 \over 3} T c_{V(SB)}(T),
\end{equation}
which are consequences of the previous relations. From these relations and their definitions in Eqs.~(A3-A5), one also has

\begin{equation}
C_{SB}(T) = c_{s(SB)}^2(T) = { 1 \over 3}, \quad \epsilon_{SB}(T)  - 3 P_{SB}(T) = 0, \quad T \rightarrow \infty.
\end{equation}
The right-hand side of the relations (A8) we call the general SB constant/limit and denote it as $(SB)$.
In many cases it is convenient to express the SB thermodynamic quantities and their derivatives in terms of this number,
which can be easily derived from the relations (A8).

\section{Analytical formulae for the GP thermodynamic quantities}

It is instructive to derive analytically all the necessary formulae for the thermodynamic quantities
using the GP pressure (6.1). Differentiating it in accordance with the definition (A1), on account of the
relation (6.2), one obtains

\begin{equation}
s(T) = {\partial P_g(T) \over \partial T} +  \Theta \left( {T_c \over T} - 1 \right)
{\partial L(T) \over \partial T}
+ \Theta \left( {T \over T_c} - 1 \right) {\partial H(T) \over \partial T},
\end{equation}
omitting here and everywhere below the subscript "GP" in accordance
with the remark made in subsection 6D. It is easy to see that the
entropy density has a jump at $T_c$,

\begin{equation}
\Delta s(T_c) = \left[ s(T > T_c) - s(T < T_c) \right]_{T \rightarrow T_c} = \left[ {\partial H(T) \over \partial T}
- {\partial L(T) \over \partial T} \right]_{T=T_c},
\end{equation}
where the difference in the right-hand-side of this equation has to be positive.

In the same way for the energy density, one obtains

\begin{eqnarray}
\epsilon(T) &=& T {\partial P_g(T) \over \partial T} - P_g(T)  +  \Theta \left( {T_c \over T} - 1 \right)
\left[ T {\partial L(T) \over \partial T} - L(T) \right] \nonumber\\
&+& \Theta \left( {T \over T_c} - 1 \right)
\left[ T {\partial H(T) \over \partial T} - H(T) \right].
\end{eqnarray}
The size of the discontinuity in the energy density (the latent heat (LH)) is

\begin{equation}
\epsilon_{LH}(T_c) = \Delta \epsilon(T_c) = \left[ \epsilon(T > T_c) -  \epsilon(T < T_c) \right]_{T \rightarrow T_c}
= T_c \left[ {\partial H(T) \over \partial T} - {\partial L(T) \over \partial T} \right]_{T=T_c},
\end{equation}
and thus it is in agreement with the discontinuity in the entropy density, since from Eqs.~(B2) and (B4)
it follows that $\epsilon_{LH}(T_c) = \Delta \epsilon(T_c) = T_c \Delta s(T_c)$.

The last independent thermodynamic quantity is the heat capacity defined in Eq.~(A2).
Differentiating the entropy density (B1), one finally obtains

\begin{eqnarray}
c_V(T) &=& T {\partial^2 P_g(T) \over \partial T^2} +  \Theta \left( {T_c \over T} - 1 \right)
 T {\partial^2 L(T) \over \partial T^2}
+ \Theta \left( {T \over T_c} - 1 \right) T {\partial^2 H(T) \over \partial T^2} \nonumber\\
&-& {T_c \over T} \delta \left( {T_c \over T} - 1 \right) {\partial L(T) \over \partial T}
+ {T \over T_c} \delta \left( {T \over T_c} - 1 \right) {\partial H(T) \over \partial T}.
\end{eqnarray}
The important observation is that the heat capacity has a $\delta$-type singularity
(an essential discontinuity) at $T=T_c$, so that the velocity of sound squared (A3) at this point is
zero, namely

\begin{equation}
c_s^2(T_c)  = { s(T_c) \over c_V(T_c)} =0.
\end{equation}
The analytical expression for the velocity of sound squared (A3) can be found with the help of
Eqs.~(B1) and (B5).

On account of Eqs.~(B3) and (6.1), the trace anomaly relation (A5) or, equivalently, the interaction measure looks like

\begin{eqnarray}
I(T) = \epsilon(T) - 3 P(T) &=& T {\partial P_g(T) \over \partial T} - 4P_g(T)  +  \Theta \left( {T_c \over T} - 1 \right)
\left[ T {\partial L(T) \over \partial T} - 4L(T) \right] \nonumber\\
&+& \Theta \left( {T \over T_c} - 1 \right)
\left[ T {\partial H(T) \over \partial T} - 4H(T) \right].
\end{eqnarray}
As mentioned above it assists in the evaluation of the temperature dependence of the gluon
condensate (A6).

The sum between the pressure (6.1) and the energy density (B3), which is nothing but the above-mentioned
enthalpy density (A7) is

\begin{equation}
e(T) = T {\partial P_g(T) \over \partial T} +  \Theta \left( {T_c \over T} - 1 \right)
T {\partial L(T) \over \partial T} + \Theta \left( {T \over T_c} - 1 \right)
T {\partial H(T) \over \partial T}.
\end{equation}

Let us point out that the discontinuities which appear in the
derivatives of the pressure are not due to the $\Theta$-functions in
Eq.~(6.1). They are due to the fact that the derivatives of the
auxiliary functions $L(T)$ and $H(T)$ are different from each other
and they are not zero at $T_c$, see Eqs.~(B2), (B4) and Eq.~(B5),
respectively. The deep reason of these discontinuities is the
principal difference between the independent basic functions
$P_{SB}(T)$ and $P_g(T)$ from each other. However, the main contributions
to the numerical values of these discontinuities come from $P_g(T)$ and its
derivatives at $T_c$, which behave rather differently below and above $T_c$
(see appendix E below). The auxiliary functions which have finally been found
within our approach in terms of the basic functions are as follows:

\begin{eqnarray}
L(T) &=&  \left[ 0.015732 e^{- 39.1 ((T_c/T)-1)} + 0.003884 e^{- 3.4 ((T_c/T)-1)} \right]P_{SB}(T)
- e^{- 0.001 (T_c/T)} P_g(T), \nonumber\\
H(T) &=& \left[ 1 - (0.22037 /t) + (0.033 \ln t / t^2) \right] P_{SB}(T) - [1.55 + 0.8482 ( T_c / T)^3] P_g(T),
\end{eqnarray}
and $t = 1 + 0.1929 \ln (T /T_c)$
as it comes out from Eqs.~(6.43) - (6.46). $P_{SB}(T)$ is given in the relations (A8), while the  NP
contribution $P_g(T)$ is given in Eq.~(5.2), and its numerical values are listed in Table I.

\section{The $\beta$-function for the confining effective charge at non-zero temperature}

Let us show explicitly the corresponding
$\beta$-function for the INP effective charge (2.6). From the renormalization group equation,

\begin{equation}
q^2 {d \alpha^{INP}(q^2; \Delta^2) \over dq^2} = \beta(\alpha^{INP}(q^2;
\Delta^2)),
\end{equation}
it simply follows that

\begin{equation}
\beta(\alpha^{INP}(q^2; \Delta^2))= - \alpha^{INP}(q^2; \Delta^2) = - {\Delta^2 \over q^2}.
\end{equation}
Thus, the corresponding $\beta$-function as a function of its
argument is always in the domain of attraction (i.e., negative). So
it has no infrared (IR) stable fixed point indeed as it is required
for the confining theory \cite{26}. Let us remind that the confining
effective charge (C2), and hence its $\beta$-function, is a result
of the summation of the skeleton (i.e., NP) loop diagrams,
contributing to the full gluon self-energy in the $q^2 \rightarrow
0$ regime (the above-mentioned cluster expansion but in powers of the mass gap).
This summation has been performed within the corresponding equations of motion
\cite{7} (and references therein).

In frequency-momentum space from Eqs.~(2.6) and (C2) one gets

\begin{equation}
\beta^{INP}( \omega^2, \omega_n^2) = - \alpha^{INP}( \omega^2, \omega_n^2) =
- { \Delta^2 \over \omega^2 + \omega_n^2}.
\end{equation}
The confining effective charge with the corresponding
$\beta$-function (C3) determines the structure of $SU(3)$ GP at low and finite
frequencies $\omega^2$ and temperatures $\omega_n^2$
(mixed phase) \cite{51} within our approach. It is worth emphasizing once more that it makes it possible to perform an
exact summation over the Matsubara frequencies in order to calculate the NP pressure $P_g(T)$ (see sections III-V).
In the limit of very high
frequencies and temperatures its role is substantially decreased, as expected, see Fig. 5 and discussion around it.

\section{The temperature dependence of $\alpha_s(T)$}

The temperature dependence of the strong fine-structure constant $\alpha_s(T)$ is determined by the renormalization group equation for the
perturbative $\beta$-function as follows \cite{12,23}:

\begin{equation}
T\frac{\partial\alpha_s}{\partial T} = -b_0\alpha^2_s - b_1\alpha^3_s  + ... = \beta(\alpha_s),
\end{equation}
i.e., the corresponding $\beta$-function is given in the two loop approximation, since the coefficients of the higher order terms $b_2, \ b_3$ are renormalization-scheme-dependent (that is  why we restrict ourself to this approximation only, and it is in agreement with accuracy
of our calculated numbers, as it was pointed out in subsections VI.A and VI.B).
The coefficients $b_0$ and $b_1$ are $b_0=11/4\pi$ and $b_1=51/8\pi^2$, respectively, for the number of flavours $n_f=0$ \cite{29}.

Keeping only the first term proportional to $\alpha^2_s$ on the right-hand-side of Eq.~(D1), one obtains an exact solution

\begin{equation}
\alpha_s(T) = \frac{A}{(1+Ab_0{\rm ln}(T/T_c))}
\end{equation}
with the integration constant $A$. Motivated by the above form of $\alpha_s(T)$ and following \cite{12,29}, we look for the solution of (D1) as

\begin{equation}
\alpha_s(T) = \frac{A}{t}+B\frac{{\rm ln}t}{t^2}\ ,\quad t = 1+Ab_0{\rm ln}(T/T_c)\ .
\end{equation}
The constants $A$ and $B$ are determined by the numerical solution of Eq.~(D1). During the numerical calculation the initial condition for the fine-structure constant was chosen as $\alpha_s(M_Z)=0.1184$ \cite{29}. The constant $A$ is determined by the value of $\alpha_s(T)$ of the numerical solution at $T=T_c$. The constant $B$ was chosen as the difference between the numerical solution and the empirical form (D3) is less than $ 10^{-3}$ in the considered temperature range below $100T_c$. These considerations were fulfilled by the values

\begin{equation}
A = 0.22037\ ,\quad B = -0.033,
\end{equation}
and the constant before the logarithmic term in $t$ becomes $Ab_0=0.1929$. Thus finally we have

\begin{equation}
\alpha_s(T) = \frac{0.22037}{t} - 0.033 \frac{{\rm ln}t}{t^2}\ ,\quad t = 1+0.1929{\rm ln}(T/T_c), \ T_c = 266.5 \ MeV.
\end{equation}
The possible structures in $\alpha_s(T)$, containing $t^{-3}, \ t^{-3} \ln t$ and $t^{-3} \ln^2 t$ combinations and depending on the coefficients $b_0$ and $b_1$, are numerically very small. For simplicity, we omit them, but it is worth pointing out that their contribution are effectively present in the numerical value of the second coefficient in Eq.~(D5).
As mentioned above this empirical solution coincides very well with the numerical solution of Eq.~(D1), see Fig. 13.

It is instructive also to explicitly present analytic expressions for the derivatives of the running effective charge $\alpha_s(T)$, namely

\begin{equation}
T \alpha'_s(T) = - \frac{0.04251}{t^2} - {0.00636 \over t^3} ( 1 -2 \ln t),
\end{equation}
and

\begin{equation}
T^2\alpha''_s(T) = \frac{0.0164}{t^3} + {0.00368 \over t^4} ( 1 -2 \ln t) + \frac{0.00245}{t^4} - T \alpha'_s(T).
\end{equation}
Let us remind that we denote any derivative with respect to the temperature $T$ in some places, for example, as follows:
$\alpha'_s(T)= (\partial \alpha_s(T) / \partial T), \ P'(T)= (\partial P(T) / \partial T)$ and so on, for convenience.

It is worth discussing now one of the important issues, namely how to approach the SB limit.
In subsection VI.A we advocate that the PT part of the gluon plasma pressure $P_{GP}(T)$
above $T_c$ can be chosen as follows: $P_{PT}(T)=f_h(T)P_{SB}(T)=(1 - \alpha_s(T))P_{SB}(T)$, where $\alpha_s(T)$ given by the empirical solution (D5).
It presents the expansion in powers of $\alpha_s=0.1184$ which can be explicitly included into the all
numerical numbers which appear in the solution (D5), and then it can be formally expand in integer powers of a small $\alpha_s$. In such kind
of the expansion there is no place for a non-analytic dependence on $\alpha_s$. That is why this expansion is convergent, i.e., any
further calculated term is always smaller than the previous one. This is also true for the initial renormalization group equation
(D1), which includes only integer powers of $\alpha_s(T)$. So one can conclude that the above-discussed composition correctly describes the AF approach
to the SB limit, i.e., it should be approached slowly (logarithmically) and always from below, not depending on the order of the PT expansion used.
Any correctly calculated NP contribution has not to change such character of the behavior of the full pressure at very high temperatures (for more detail discussion go back to subsection VII.A, in particular see Fig. 5 and discussion around it).

However, the direct analytical derivations in the thermal PT QCD have discovered a non-analytic dependence
on the coupling constant in the PT series for the pressure (see, for example, \cite{22,50} and references therein).
So the PT QCD pressure (normalized to the SB pressure and possessing correct high temperature limit) in this case can be expressed as follows:

\begin{equation}
P_{PT}(T)=f_h(T)P_{SB}(T)=(1 - f_s(T))P_{SB}(T),
\end{equation}
where

\begin{equation}
f_s(T) = a_1 \cdot \alpha_s(T) + a_2 \cdot \alpha^{3/2}_s(T) + a_3 \cdot \alpha^2_s(T) + a_4 \cdot \alpha^2_s(T) \cdot \log \alpha_s(T) +
a_5 \cdot \alpha^{5/2}_s(T) + a_6 \cdot \alpha^3_s(T) + a_7 \cdot \alpha^3_s(T) \cdot \log \alpha_s(T) + ...
\end{equation}

For the numerical values of the coefficients $a_i$ (including their signs) and
analytic expression for $\alpha_s(T)$ itself see \cite{37}. A non-analytic dependence of the PT pressure on the temperature-dependent
coupling constant means that the above-shown expansion is not convergent. For example, if somebody will be able to analytically derive
the term beyond the last analytically known $\alpha^3 \log \alpha$-order term, nevertheless, it may be positive and numerically much bigger than
the previous one. Such an effect is clearly seen in Fig. 14, where the $g^3$-order term is numerically bigger than the previous one, indeed.
In this case, the SB limit will be approached from above, which is not acceptable, of course. In other words, the expansion for $\alpha_s(T)$
in powers of $\alpha_s$ is convergent, while the expansion (D9) in powers of $\alpha_s(T)$ itself is not.

However, up to the $\alpha^3 \log \alpha$-order term the pure gauge lattice QCD \cite{37} (and references therein) is compatible with
analytic one as it follows from Fig. 14 as well. So we are going to describe the lattice results
for the pure gauge QCD obtained in \cite{37} in a forthcoming paper. In any case, their simulations need an independent investigation within our
general formalism, since they fix the SB limit explicitly, while in the pure gauge lattice QCD simulations \cite{33} (used in the present paper)
this limit has not been fixed analytically. Also the value of the lattice pressure \cite{37} at $T_c$ is rather different from that in \cite{33},
though the value of $T_c$ itself is almost the same in both simulations.

\begin{figure}
\begin{center}
\includegraphics[width=10cm]{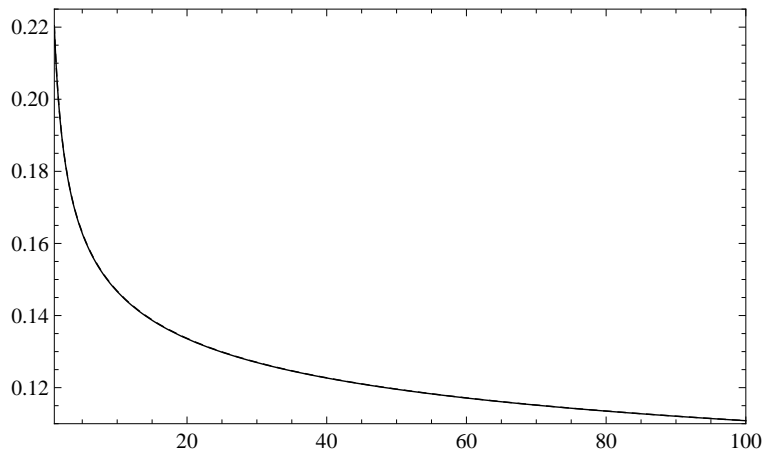}
\caption{Numerical solution of Eq.~(D1) is shown as solid curve. The empirical solution Eq.~(D5) is shown as dashed curve.
They completely coincide up to third digit after point.}
\label{fig:1}
\end{center}
\end{figure}

\begin{figure}
\begin{center}
\includegraphics[width=10cm]{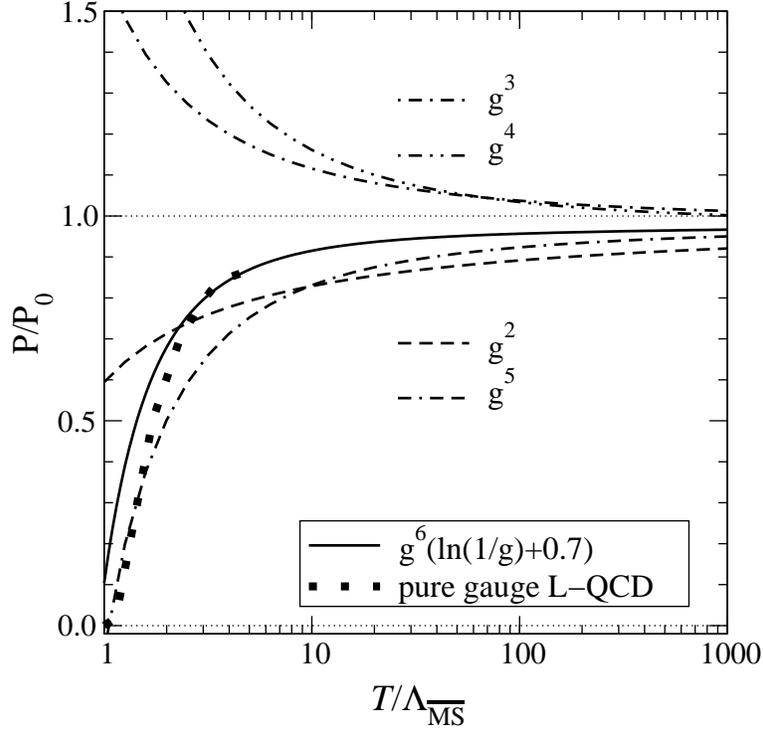}
\caption{Non-analytic PT QCD pressure normalized to the SB pressure denoted as $P_0$, results of \cite{50}, but the picture itself
is taken from \cite{12}.}
\label{fig:3}
\end{center}
\end{figure}

\section{Analytical and numerical evaluation of the latent heat}

It is instructive to calculate the auxiliary functions $L(T)$ and
$H(T)$ starting from their general expressions (6.11) and (6.12). Of
course, the final numerical results are consistent with their
derived expressions shown in Eq.~(B9). From the expression (6.11) it
follows that

\begin{eqnarray}
{\partial L(T) \over \partial T} &=& { T_c \over T^2}
\sum_{i=1}^n \mu_i A_i e^{- \mu_i (T_c / T)} P_{SB}(T) + \sum_{i=1}^n A_i e^{- \mu_i (T_c / T)}
{\partial P_{SB}(T) \over \partial T} \nonumber\\
&-& { T_c \over T^2} \sum_{i=n+1}^m \mu_i A_i
e^{- \mu_i (T_c / T)} P_g(T) - \sum_{i=n+1}^m A_i e^{- \mu_i (T_c / T)} {\partial P_g(T) \over \partial T},
\end{eqnarray}
and at $T=T_c$ it is

\begin{eqnarray}
\left( {\partial L(T) \over \partial T} \right)_{T_c} &=&
{ 1 \over T_c } \sum_{i=1}^n \mu_i A_i e^{- \mu_i} P_{SB}(T_c) + \sum_{i=1}^n A_i e^{- \mu_i}
\left( {\partial P_{SB}(T) \over \partial T} \right)_{T_c} \nonumber\\
&-& { 1 \over T_c } \sum_{i=n+1}^m \mu_i A_i e^{- \mu_i} P_g(T_c) -
\sum_{i=n+1}^m A_i e^{- \mu_i} \left( {\partial P_g(T) \over \partial T} \right)_{T_c}.
\end{eqnarray}

In the same way from the expressions (6.12) and (6.27) it follows that

\begin{equation}
{\partial H(T) \over \partial T} = ( 1 - \alpha_s(T))  {\partial P_{SB}(T) \over \partial T}
- \alpha'_s(T)P_{SB}(T) + { n \nu \over T_c } \left({ T \over T_c} \right)^{-n-1} P_g(T)
- \phi_h(T){\partial P_g(T) \over \partial T},
\end{equation}
and at $T=T_c$ it is

\begin{equation}
\left( {\partial H(T) \over \partial T} \right)_{T_c} = 0.77963 \left(
{\partial P_{SB}(T) \over \partial T} \right)_{T_c} + {0.04887 \over T_c}P_{SB}(T_c)
+ { n \nu \over T_c} P_g(T_c) - \phi_2(T_c) \left( {\partial P_g(T) \over
\partial T} \right)_{T_c}.
\end{equation}

Taking further into account these relations and the relations (6.4) and (6.14), the latent
heat (B4) thus becomes

\begin{equation}
\epsilon_{LH}(T_c) = \left[ n \nu  - 1.839855 \left( \sum_{i=1}^n \mu_i A_i
e^{- \mu_i} - 0.04887 \right) + \sum_{i=n+1}^m \mu_i A_i e^{- \mu_i}   \right] P_g(T_c),
\end{equation}
and in dimensionless units it is

\begin{equation}
\epsilon_{LH}  \equiv { \epsilon_{LH}(T_c) \over T^4_c} =
\left[ n \nu + 0.0899 - 1.839855 \sum_{i=1}^n \mu_i A_i e^{- \mu_i} + \sum_{i=n+1}^m \mu_i A_i e^{- \mu_i} \right] \times 0.95366,
\end{equation}
where the number $P_g(T_c)/T_c^4$ taken from Table I has been already substituted. Remembering that in the sum over $i$ the number $n=2$
and the number $m=3$, and using further the numbers (6.34) and (6.41), one finally obtains

\begin{equation}
\epsilon_{LH} = 1.41,
\end{equation}
in complete agreement with lattice data discussed in subsection VII.D. However, let us emphasize that
we did not use the lattice data for the energy density as well as for all other thermodynamic quantities, only for the pressure.

\section{Least Mean Squares method and the definition of the average deviation}

In order to adjust lattice data at high temperatures above
$T_c$ we use Eq. (6.33), which depends on the parameters
$\nu$ and $n$. For a given $n$ the parameter $\nu$ is determined by
using the LMS method \cite{41}. This method makes it possible to calculate the
values of the parameter $\nu$ for the last $p\geq 2$ number of data
points available from lattice results. In this way the temperature
region below $T/T_c = 3.436657$ up to $T/T_c = 1$ was covered with an approximation curve
which best fits to lattice calculations according to the LMS method.
Let us note that for the accuracy of this method the number of data points $p$ has to be sufficiently big.

However, the result of the calculations for $\nu$ depends on the
number of data points $p$ considered and the value for $n$, i.e.,
$\nu=\nu(p,n)$. In order to choose which value for $n$ is
preferable, we have introduced the average deviation $\Delta_p(n)$
for the given numbers of points $p=162$. For our purpose it is
convenient to define this quantity as follows:

\begin{equation}
\Delta_p(n) = \frac{1}{p}\sum_{i=1}^p \frac{1}{T_i^4}\left[ (SB)\,
P_l(T_i) - 3P_{GP}(T_i,n) \right]^2 \ .
\end{equation}
Here $T_i$ denotes the temperature at the $i^{th}$ data point, while $(SB)$ is the above-mentioned SB general constant.
As a result of the numerical investigations we have found that the LMS method gives the best fit for $n=3$. The average deviation is minimal
at a sufficiently large number of lattice data points $p_{min}=114$, see Fig. 15. This makes it possible to finally fix $\nu=0.8482$ and thus $\nu_0=0.55$
via Eq.~(6.32).

\begin{figure}
\begin{center}
\includegraphics[width=10cm]{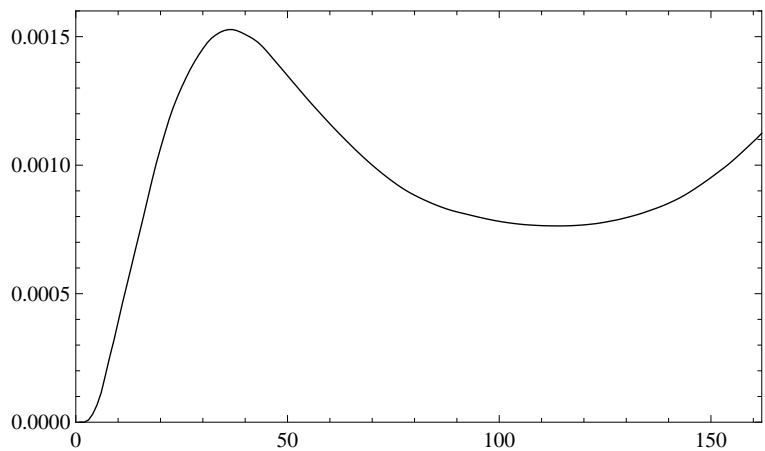}
\caption{Average deviations $\Delta_p(n)$ as functions of lattice data points $p$ for $n=3$.}
\label{fig:1}
\end{center}
\end{figure}

\begin{table}[H]
\centering
\caption{The gluon pressure (5.2)
versus the GP pressure (6.43)}\label{table: nonlin}
\vspace {5mm}
\begin{tabular}{| l|l |l |l| l| }
\hline
 $T/T_{c}$ & $3P_{g}(T)/T^{4}$& $3P_{NP}(T)/T^{4}$ & $3P^{s}_{PT}(T)/T^{4}$ & $3P(T)/T^{4} $\\ [0.5ex]
 \hline
 & & & & \\
  0.1 & $2.2569 \cdot {10}^{-7}$ & $1.9915 \cdot {10}^{-7}$ & $ 2.6541 \cdot {10}^{-8}$ & $2.2558 \cdot {10}^{-10}$ \\
 0.2 & 0.00534166 & 0.00448888 & 0.000852786 & $2.6955 \cdot {10}^{-6}$ \\
 0.3 & 0.121856 & 0.104807 & 0.0170489 & 0.0000479432 \\
 0.4 & 0.520086 & 0.458813 & 0.0612728 & 0.00025465 \\
 0.5 & 1.13612 & 1.01973 & 0.116396 & 0.000909504 \\
 0.6 & 1.77261 & 1.60796 & 0.164649 & 0.00241463 \\
 0.7 & 2.28665 & 2.08719 & 0.199454 & 0.0050882 \\
 0.8 & 2.62868 & 2.40758 & 0.221099 & 0.00907158 \\
 0.9 & 2.80883 & 2.57659 & 0.232237 & 0.0153992 \\
 1 & 2.86098 & 2.62523 & 0.235748 & 0.103562 \\
 1.1 & 2.82211 & 2.58807 & 0.234034 & 0.77721 \\
 1.2 & 2.72373 & 2.49483 & 0.228899 & 1.31381 \\
 1.3 & 2.58982 & 2.36819 & 0.221634 & 1.7432 \\
 1.4 & 2.43743 & 2.22429 & 0.213135 & 2.09012 \\
 1.5 & 2.2781 & 2.07409 & 0.204011 & 2.37372 \\
 1.6 & 2.11933 & 1.92466 & 0.194673 & 2.60841 \\
 1.7 & 1.96576 & 1.78037 & 0.185387 & 2.8049 \\
 1.8 & 1.82007 & 1.64375 & 0.176326 & 2.97124 \\
 1.9 & 1.68366 & 1.51606 & 0.167595 & 3.11344 \\
 2 & 1.55705 & 1.3978 & 0.159256 & 3.23611 \\
 2.1 & 1.44026 & 1.28892 & 0.151339 & 3.34277 \\
 2.2 & 1.33295 & 1.18909 & 0.143854 & 3.43617 \\
 2.3 & 1.23461 & 1.09781 & 0.136798 & 3.51849 \\
 2.4 & 1.14464 & 1.01448 & 0.130159 & 3.59145 \\
 2.5 & 1.0624 & 0.938478 & 0.12392 & 3.65647 \\
 2.6 & 0.987234 & 0.869172 & 0.118063 & 3.71467 \\
 2.7 & 0.918531 & 0.805967 & 0.112564 & 3.76699 \\
 2.8 & 0.855703 & 0.7483 & 0.107403 & 3.81422 \\
 2.9 & 0.798208 & 0.695651 & 0.102557 & 3.85701 \\
 3 & 0.745547 & 0.64754 & 0.0980066 & 3.89591 \\
 3.1 & 0.697264 & 0.603533 & 0.0937309 & 3.93138 \\
 3.2 & 0.652946 & 0.563235 & 0.0897111 & 3.96381 \\
 3.3 & 0.61222 & 0.526291 & 0.0859295 & 3.99357 \\
 3.4 & 0.574749 & 0.49238 & 0.0823693 & 4.02092 \\
 3.5 & 0.54023 & 0.461215 & 0.0790152 & 4.04614 \\
 4 & 0.403268 & 0.338401 & 0.0648672 & 4.14704 \\
 4.5 & 0.309 & 0.254884 & 0.054116 & 4.21843 \\
 5 & 0.242188 & 0.196402 & 0.045786 & 4.27112 \\
 6 & 0.157364 & 0.123414 & 0.0339498 & 4.34306 \\
 7 & 0.108525 & 0.0823814 & 0.0261433 & 4.3897 \\
 8 & 0.0783762 & 0.0576392 & 0.0207371 & 4.4225 \\
 9 & 0.0587111 & 0.0418673 & 0.0168438 & 4.44702 \\
 10 & 0.0453013 & 0.031352 & 0.0139493 & 4.46623 \\
 20 & 0.00827882 & 0.00442155 & 0.00385727 & 4.5559 \\
 30 & 0.00313823 & 0.00136433 & 0.0017739 & 4.59358 \\
 40 & 0.00160267 & 0.000587514 & 0.00101515 & 4.61705 \\
 50 & 0.000961058 & 0.000304574 & 0.000656484 & 4.63389 \\
 60 & 0.000636814 & 0.00017774 & 0.000459074 & 4.6469 \\
 70 & 0.00045157 & 0.000112607 & 0.000338964 & 4.65743 \\
 80 & 0.000336275 & 0.0000757821 & 0.000260493 & 4.66624 \\
 90 & 0.000259838 & 0.0000534144 & 0.000206423 & 4.67378 \\
 100 & 0.000206646 & 0.000039051 & 0.000167595 & 4.68036 \\ [1ex]
 \hline
\end{tabular}
\end{table}

\end{document}